\documentclass[12pt, onecolumn]{article}

\usepackage{lineno}
\usepackage{url}

\usepackage{graphicx}
\usepackage{epsfig} 
\usepackage{amssymb, amsmath, mathrsfs}
\usepackage{float}
\usepackage{verbatim} 
\usepackage{caption}
\usepackage{multirow}
\usepackage{subfig}
\usepackage[colorlinks]{hyperref}
\usepackage[a4paper, margin=1in]{geometry}

\allowdisplaybreaks

\newcommand{\Stil}{\tilde{S}}
\newcommand{\Itil}{\tilde{I}}
\newcommand{\Rtil}{\tilde{R}}
\newcommand{\Ftil}{\tilde{F}}
\newcommand{\xtil}{\tilde{x}}
\newcommand{\xtilb}{\mathbf{\xtil}}

\newcommand{\undc}{\underline{c}}

\newcommand{\hessinv}{\nabla^{-2}}
\newcommand{\pa}{\partial}
\newcommand{\thetab}{\pmb{\theta}}
\newcommand{\lambdab}{\pmb{\lambda}}
\newcommand{\phib}{\pmb{\phi}}
\newcommand{\xb}{\mathbf{x}}
\newcommand{\tb}{\mathbf{t}}
\newcommand{\yb}{\mathbf{y}}
\newcommand{\fb}{\mathbf{f}}
\newcommand{\cb}{\mathbf{c}}
\newcommand{\deldelt}{\frac{\pa}{\pa t}}
\newcommand{\nphi}{\nabla_{\pmb{\phi}}}
\newcommand\numberthis{\addtocounter{equation}{1}\tag{\theequation}}

\begin{document}
	
	\begin{center}{\large Deep Learning Aided Laplace Based Bayesian Inference for Epidemiological Systems}
		\end{center}

	\begin{center}{\small
		Wai M. Kwok\textsuperscript{1},
		George Streftaris\textsuperscript{2},
		Sarat C. Dass\textsuperscript{1*}
	}
	\end{center}
	
	\bigskip
	{\small \textbf{1} School of Mathematical and Computer Sciences, Heriot-Watt University Malaysia, 62200 Putrajaya, Malaysia
	\\
	\textbf{2} Mathematical and Computer Sciences, Heriot-Watt University, Edinburgh, United Kingdom, Maxwell
	Institute for Mathematical Sciences, Edinburgh, United Kingdom}
	\bigskip 
	\\
	{\small*Corresponding author, S.Dass@hw.ac.uk}

	\begin{abstract}
		\normalsize		
		\noindent 
		Parameter estimation and associated uncertainty quantification is an important problem in dynamical systems characterised by ordinary differential equation (ODE) models that are often nonlinear. Typically, such models have analytically intractable trajectories which result in likelihoods and posterior distributions that are similarly intractable. Bayesian inference for ODE systems via simulation methods require numerical approximations to produce inference with high accuracy at a cost of heavy computational power and slow convergence. At the same time, Artificial Neural Networks (ANN) offer tractability that can be utilized to construct an approximate but tractable likelihood and posterior distribution. In this paper we propose a hybrid approach, where Laplace-based Bayesian inference is combined with an ANN architecture for obtaining approximations to the ODE trajectories as a function of the unknown initial values and system parameters. Suitable choices of a collocation grid and customized loss functions are proposed to fine tune the ODE trajectories and Laplace approximation. The effectiveness of our proposed methods is demonstrated using an epidemiological system with non-analytical solutions – the Susceptible-Infectious-Removed (SIR) model for infectious diseases – based on simulated and real-life influenza datasets. The novelty and attractiveness of our proposed approach include (i) a new development of Bayesian inference using ANN architectures for ODE based dynamical systems, and (ii) a computationally fast posterior inference by avoiding convergence issues of benchmark Markov Chain Monte Carlo methods. These two features establish the developed approach as an accurate alternative to traditional Bayesian computational methods, with improved computational cost.
	\end{abstract}
	
	\begin{center}
			Neural Network; Forward Problems; Laplace Approximation; Bayesian Inference; \\ Parameter Estimation; Inverse Problems; Epidemics.
	\end{center}	

\nocite{*}

\section{Introduction}
\label{section:intro}

\subsection{Existing Works and Literature}
Ordinary differential equation (ODE) models refer to statistical models whose mean functions are governed by an underlying ordinary differential equation. Dynamical systems are systems that describe changes in certain states or quantities over time, and are readily formalized by ordinary differential equations. As a result, dynamical systems elicited via ODE models find applications in numerous fields including biology, epidemiology, and physics, indicating their importance in these fields for understanding and gaining insights into the processes involved. An important aspect of these application problems is the inference of unknown parameters of ODE models and associated uncertainty quantification. This is known to be a very challenging problem due to the intractability of the underlying ODE, which is typically nonlinear and thus prohibits any closed form solution. Bayesian inference of unknown parameters of ODE models is also challenging due to this intractability which result in likelihoods and posterior distributions that are similarly intractable. Thus, Bayesian computations for inference in ODE models are carried out via simulation methods that require numerical approximations to produce inference with high accuracy at a cost of heavier computational power and slow convergence. Examples of Bayesian inference for ODE models depending on long-run convergence results of iterative procedures are reported in \cite{roda2020bayesian, chen2012monte, christensen2001bayesian}, and associated convergence diagnostic methods can be found in \cite{gelman1992inference, brooks1998general}.\\

Recent advances in computational power have brought about increasingly complex models and innovative solutions to both traditional and new problems. One such example is the use of artificial neural networks (ANNs) in mathematical modelling. ANNs were first introduced in \cite{mcculloch1943logical}, and research into their theoretical capabilities \cite{cybenko1989approximation, hornik1989multilayer} laid strong foundations for the usage of neural networks as a universal function approximation tool \cite{adcock2021gap, zainuddin2008function, yang2013investigation}; Li (1996) \cite{li1996simultaneous} further showed that multivariate functions and their derivatives can be simultaneously approximated by ANNs, with a comprehensive, practical algorithm that improves the optimization of ANN weights and architectures laid out by \cite{nguyen1999approximation} as well as references therein. Crucially, the optimal or example ANNs in these literature often use less than 3 hidden layers and less than 20 neurons in each layer to achieve good results, which can be noted to be relatively simple architectures to achieve effective functional and derivative approximations. Since then, ANN training algorithms that explicitly incorporate derivative information have been explored in an effort to enhance their approximation abilities \cite{avrutskiy2020enhancing, pukrittayakamee2011practical, ferrari2005smooth, basson1999approximation}. \\

The traditional way of approximating solutions and trajectories of nonlinear ODEs are based on numerical methods such as the Euler, Runge-Kutta, finite difference and finite element methods, when closed analytic-form solutions are unavailable. Nevertheless, noting that realized states of the ODE are simply functions of time and intrinsic parameters (including the unknown initial values), the idea of solving differential equations, or at least approximating their dynamics, with ANNs seems natural. This is the approach taken in this paper where we approximate the functional value and derivatives of the underlying ODE in dynamical systems by an appropriate ANN. The resulting ANN is tractable and allows derivatives of the likelihood and posterior densities to be computed in closed form. As a result, Bayesian inference is facilitated in closed form based on the Laplace approximation to the posterior distribution of the unknown parameters. The effectiveness of our proposed hybrid Bayesian-ANN methods as an alternative to traditional iterative Bayesian methods like MCMC is demonstrated using an epidemiological system with non-analytical solutions – the Susceptible-Infectious-Removed (SIR) model for infectious diseases – based on simulated and real-life influenza datasets. \\

The approximation by ANNs of ODE models is not new. Physics-Informed Neural Networks (PINNs) proposed by Raissi \textit{et al} (2019) \cite{raissi2019physics} are unsupervised neural networks (i.e. no training targets) trained to obey the physics of the dynamical system by using penalties based on the differential equations in the loss function. Raissi \textit{et al} \cite{raissi2019physics} also develops methodology for data discovery, i.e. parameter learning, based on a given observed trajectory of the ODE system for unknown parameters. Their approach is similar to the approach of Ramsay (2007) \cite{ramsay2007parameter} where in the latter, a parameter cascading approach is implemented to estimate the parameters of the ODE as oppposed to the joint minimization approach of Raissi \textit{et al} \cite{raissi2019physics}. Ramsay \cite{ramsay2007parameter} also incorporates a penalty coefficient and selects it in a data driven way. Raissi \textit{et al} \cite{raissi2019physics} does not incorporate a penalty coefficient and their parameter estimates do not come with associated measures of uncertainty. \\

Further work on approximating ODEs by extension of ANNs have been reported in the literature. A deep neural network approach to forward-inverse problems by Jo \textit{et al} \cite{jo2020deep} extends results from Li \cite{li1996simultaneous} to prove that under certain conditions, these DNN approximations converge to the dynamical system's solution, and the parameter estimates resulting from the ANN converge to their true values. Jo \textit{et al} subsequently applied this methodology to a Susceptible-Infected-Removed (SIR) model for COVID-19 cases in South Korea \cite{jo2020analysis}. Their joint minimization approach is similar to that of Raissi \textit{et al} \cite{raissi2019physics} for parameter learning and they, too, do not report uncertainty measures corresponding to their parameter estimates. Another drawback of the theoretical approach of Jo \textit{et al} \cite{jo2020deep} is that a "close-enough" ANN is shown to exist but it does not indicate how well the learned network estimates the ODE system when the loss function has been trained to be less than $\epsilon$, for some small $\epsilon > 0$. The DNNs in these works use 4 or more layers, with over 100 neurons in each layer, demonstrating the high complexity of these models.

\subsection{Our Contributions}
Previous works using ANNs for ODE models do not incorporate uncertainly measures (either summary measures or entire distributions) associated with the estimates of unknown parameters of the ODE models. In this paper, we develop a Bayesian inferential framework for analytically intractable ODE models by developing ANN architecture that tractably approximates the function values and derivatives of the ODE.  Uncertainty measures are obtained by additionally deriving a Laplace approximation to the true posterior distribution, which is a Gaussian distribution whose variance relies on the ANNs' approximations to the ODE system's partial derivatives. The proposed hybrid method admits an approximate but tractable posterior distribution, and thereby enables us to obtain accurate parameter point estimates together with appropriate uncertainty quantifications. \\

Our approach to the ANN approximations differs from Raissi \textit{et al} \cite{raissi2019physics}, Jo \textit{et al} \cite{jo2020deep} and Ramsay \cite{ramsay2007parameter} in that we utilize \textit{supervised} ANNs with the unknown ODE parameters as extrinsic \textit{inputs}, trained by numerical solutions of the ODE over a grid of collocation points in the parameter and temporal domains; hence the inclusion of initial values as an input further allows initial value estimation. In inference for epidemic systems, this is particularly important as it allows the estimation of crucial epidemic quantities, such as the number of infected individuals at the onset of an outbreak. Under such an architecture, obtaining the partial derivatives of the ODE system's states with respect to its parameters (and initial values) is straightforward. This facilitates the imposition of regularizations on the ANNs' loss functions to improve their accuracies in approximating the solutions and partial derivatives of the ODE, and thus provides more reliable estimates of posterior variances in the aforementioned Laplace approximation procedure.\\

A further important contribution of this work is that the hybrid method combining Bayesian inference with an appropriately regularized ANN architecture proves to be a faster and more lightweight alternative to typical MCMC methods that require long runs and convergence of iterative chains. The latter are well-known methods in the Bayesian framework for obtaining inference with associated measures of uncertainty. We demonstrate the gain in computational costs in our simulation and real data examples subsequently. 

\subsection{Organization of sections}

The remainder of the paper is organized as follows: Section \ref{section:prelim} explains the some preliminary concepts leading up to our methods, and Section \ref{section:methods} details the application of said methods using an example dynamical system - the SIR Model, on simulated datasets. Subsequently, Section \ref{section:influenza} demonstrates the usage of our methods to infer epidemic parameters from a real-life influenza dataset. Section \ref{section:discussion} discusses the results obtained in Sections \ref{section:methods} and \ref{section:influenza}, and reflects on the advantages and limitations of our methods. Finally, conclusions and potential future work are outlined in Section \ref{section:conclusion}.

\section{Methods}
\label{section:prelim}

\subsection{ODE Systems}
\label{subs:introbayes}

Differential equations arise in many areas and are often used to mathematically describe rates of changes of a system's states, with respect to some underlying variable (such as time). Specifically, let $\xb(t) := \left(x_1(t), x_2(t), ..., x_D(t)\right) \in \mathbb{R}^D$ denote the states of an ODE system whose values vary over time points $t \in [T_0, T_1]$ based on the differential equation, where $T_0$ and $T_1$ denote the initial and final time points respectively, The general ODE equation can be written as:
\begin{align*}
	\frac{d\xb(t)}{dt} = \fb(\xb(t), t, \thetab)
	\numberthis
	\label{eq:dynsys}
\end{align*}
Here, $\fb(\xb(t), t, \thetab) = \left( f_1(\xb(t), t, \thetab), f_2(\xb(t), t, \thetab), ..., f_D(\xb(t), t, \thetab) \right) \in \mathbb{R}^D$ and $\thetab = \left(\theta_1, \theta_2, ..., \theta_Q\right) \in \mathbb{R}^Q$. This system is called autonomous if $\fb$ is further independent of $t$; in this paper, we will consider autonomous ODE systems. \\ 

The ODE system in Equation (\ref{eq:dynsys}) is completely determined based on an initial set of values $\xb(T_0) = \left( x_1(T_0), x_2(T_0), ..., x_d(T_0)\right)$ and fixed $\thetab \in \mathbb{R}^Q$, hence we often wish to draw inference on $\xb(T_0)$ alongside $\thetab$ based on observed data. It is then natural to treat $\xb(T_0)$ as an additional parameter of the ODE and append it to $\thetab$, that is, we may define $\phib := (\xb(T_0), \thetab) = \big(x_1(T_0), ..., x_D(T_0), \theta_1, ..., \theta_Q \big) \in \mathbb{R}^{D+Q}$ and seek to infer $\phib$ from observed data. The solution, or trajectory, of the system may then be represented as $\xb(t, \phib)$. \\

Additionally, in our proposed methods detailed later (pertaining to the neural network design in Section~\ref{subs:introdnn} and loss function regularizations in Section~\ref{subs:lossfn}), we will also consider the derivatives of the system's states with respect to its parameters (and initial values) $\nphi \xb(t, \phib)$:
\begin{align*}
	\nabla_\phi \xb(t, \phib) := \begin{pmatrix}
		\frac{\pa}{\pa \phi_1} x_1(t, \phib) &
		\frac{\pa}{\pa \phi_2} x_1(t, \phib) & \cdots &
		\frac{\pa}{\pa \phi_{D+Q} }x_1(t, \phib) \\
		\frac{\pa}{\pa \phi_1} x_2(t, \phib) &
		\frac{\pa}{\pa \phi_2} x_2(t, \phib) & \cdots &
		\frac{\pa}{\pa \phi_{D+Q}} x_2(t, \phib) \\
		\vdots & \vdots & \ddots & \vdots \\
		\frac{\pa}{\pa \phi_1} x_D(t, \phib) &
		\frac{\pa}{\pa \phi_2} x_D(t, \phib) & \cdots &
		\frac{\pa}{\pa \phi_{D+Q}} x_D(t, \phib)
	\end{pmatrix}
\end{align*}
whose trajectory is given by
\begin{align*}
	\frac{\pa }{\pa t}\nphi \xb(t, \phib) = \nphi \fb (\xb(t, \phib), t, \phib)
	\numberthis
	\label{eq:dynsysderiv}
\end{align*}
obtained by differentiating Equation (\ref{eq:dynsys}) with respect to $\phib$. Extending the system states $\xb$ to include these derivatives as additional states of the dynamical system, we define 
\begin{align*}
	\xb^\dagger &  = \left(x_1^\dagger, ..., x_D^\dagger, x_{D+1}^\dagger, ..., x_{D(D+Q+1)}^\dagger\right) \\ 
	& := \left(x_1, ..., x_D, \frac{\pa x_1}{\pa \phi_1}, ..., \frac{\pa x_1}{\pa \phi_{D+Q}}, \frac{\pa x_2}{\pa \phi_1}, ..., \frac{\pa x_2}{\pa \phi_D}, ..., \frac{\pa x_D}{\pa \phi_1}, ..., \frac{\pa x_D}{\pa \phi_{D+Q}}\right) \in \mathbb{R}^{D^\dagger} 
	\numberthis
	\label{eq:xdagger}
\end{align*}
where $D^\dagger := D(D+Q+1)$. Combining Equations (\ref{eq:dynsys}) and (\ref{eq:dynsysderiv}) results in a system for $\xb^\dagger$ written as
\begin{align*}
	\frac{\pa \xb^\dagger(t, \phib)}{\pa t} = \fb^\dagger(\xb^\dagger(t, \phib), t, \phib).
	\numberthis
	\label{eq:dynsysextended}
\end{align*}

\subsection{Bayesian Inference}
In a dynamical system, the observed data $y$ at different times $t$ usually arise as a result of a combination of entries of $\xb(t), \phib$ and noise terms, thus we may write the observation at time $t$ as $y(t) := y(\xb(t), \phib)$, and the associated probability as
\begin{align*}
	p(y(t_k) | \xb(t_k), \phib)
	\numberthis
	\label{eq:generallikelihoodeach}
\end{align*}
where $T_0 \le t_1 < t_2 < \cdots < t_N \le T_1$ are the time points of the observations. Assuming that the observations are independent of each other, the complete likelihood
\begin{align*}
	p(\mathbf{y}|\xb(\mathbf{t}), \phib) \equiv \prod_{k = 1}^{N} 
	p(y(t_k) | \xb(t_k), \phib)
	\numberthis
	\label{eq:generallikelihood}
\end{align*}
can be obtained by multiplying together the probabilities in (\ref{eq:generallikelihoodeach}). It is often understood that the observations $\yb$ arise from the underlying states $\xb$, in which case we may abbreviate the complete likelihood as $p(\yb|\phib)$.\\

Bayesian inference generally involves specifying a prior belief on a set of parameters, and then factoring in the likelihood of observed data to arrive at a posterior belief of said parameters. Probability distributions are often used to represent these beliefs: denote by $p(\phib)$ the prior distribution of $\phib$, and $p(\mathbf{y}|\phib)$ the likelihood of observing some data $\yb$ given the parameters $\phib$. The posterior distribution of $\phib$, $p(\phib|\yb)$, then satisfies
\begin{align*}
	p(\phib|\yb) = \frac{p(\phib) p(\yb|\phib)}{\int p(\phib) p(\yb|\phib) \ d\phib} \propto p(\phib) p(\yb|\phib),
	\numberthis
	\label{eq:bayesrule}
\end{align*}
where in the last expression, it suffices to consider the product of the likelihood and prior and view it as a function of $\phib$ only. \\

Equation (\ref{eq:bayesrule}) is the usual expression of the posterior density associated with $\phib$ given observed data. The typical approach in Bayesian inference, when the posterior is intractable, is to obtain samples from this posterior and perform Monte Carlo inference in order to derive estimates such as the posterior mean, variance and credible intervals. However, in the case of posteriors from ODE models, obtaining samples directly from them is difficult. This is because many ODE models cannot be solved analytically in closed form, resulting in a likelihood expression and hence a posterior density that is intractable. Previous works have worked around this intractability in a number of ways, such as Monte Carlo simulations with importance sampling arising from likelihoods calculated using numerical solutions \cite{dass2021data}, Approximate Bayesian Computation which avoids the use of likelihoods \cite{tavare1997inferring, jiang2017learning}, modifications to Markov Chain Monte Carlo methods \cite{middleton2020unbiased} and many more. Most of these methods are computationally slow and are either frequentist or do not provide exact Bayesian inference. \\

\subsection{Challenges of Analytically Intractable Likelihoods}
\label{subs:challenge}

The solution of a dynamical system refers to the set of values of the system's states that satisfy the corresponding ODEs; some systems of differential equations have exact, analytically tractable solutions, whereas others generally require function or numerical approximations to their true solutions. Function approximations involve pre-defining a class of functions and finding an element or a subset of the class that best match the analytical solution, and on the other hand, numerical methods often involve discretizing some parameter space and then obtaining numerical values of the system's states at those discrete points. \\

Numerical methods can prove to be inconvenient in computing the likelihoods and posterior distributions of ODE parameters: for \textit{each} $\phib_m$ (where $m \in \{1, 2, ..., M\}$), the ODE needs to be solved numerically to obtain the solution $(\xb(t_1, \phib_m), \xb(t_2, \phib_m), ..., \xb(t_N, \phib_m))$. This is often repeated for a large number of times $M$ to obtain a good representation of the posterior distribution, rendering the method computationally expensive. Any changes to the true values of the parameters $\phib$ also causes the previously-inferred posterior to be inaccurate, and the algorithm will have to be repeated for another $M$ interations to obtain new posteriors. \\

With function approximation methods, the accuracy of the solution naturally depends on the function space considered. Given selected functions $\xtilb(t, \phib) = (\xtil_1(t, \phib), \xtil_2(t, \phib), ..., \\ \xtil_D(t, \phib))$ that approximate each of $\xb(t, \phib) = (x_1(t, \phib), x_2(t, \phib), ..., x_D(t, \phib))$ in Equation (\ref{eq:dynsys}) reasonably well, this method bears the advantage that the state values $\xb$ for different $\phib_m$ may be computed simply by changing the inputs to the approximating function. Furthermore, as the function space considered is arbitrary, we are somewhat free to choose $\xtil_d$ with desirable properties such as continuity and differentiability - this proves to be useful in quantifying the posterior uncertainty in $\phib$ in later sections. \\

Next, we will employ Artificial Neural Networks (ANNs) to approximate the solution of an ODE system parameterized by $\phib$. Our proposed method utilizes ANNs as a universal function approximator for the ODE solution $\xtilb(t, \phib) \approxeq \xb(t, \phib)$, taking it explicitly as a function of $t$ and $\phib$ - we shall henceforth refer to the approximation of $\xb$ as Method I. Additionally we also consider an ANN that approximates $\xb^\dagger$, i.e. the system states as well as its first derivatives, which we shall refer to as Method II.\\

For each method, we will first describe the process of obtaining training features and targets using collocation points, then describe the architectures used, and finally discuss the custom loss functions and how they serve to regularize the ANN outputs.

\subsection{Artificial Neural Networks: Training Examples}
\label{subs:introtraining}

Consider a dynamical system that occurs over the range of time $[T_0, T_1]$ and the range of parameter values $[\phi_i^L, \phi_i^U], i = 1, ..., D+Q$. We choose collocation points $(t^C, \phib^C) = (t^C, \phi_1^C, \phi_2^C, ..., \phi_{D+Q}^C)$ in the $(t, \phib)$ domain in the following way:
\begin{align*}
	& t^C \in \{t_1^C, t_2^C, ..., t_{N_t}^C\}; \\
	& \phi_i^C \in \{\phi_{i,1}^C, \phi_{i,2}^C, ..., \phi_{i,N_{\phi_i}}^C \} \ \ \text{for } i = 1, 2, ..., D+Q,
\end{align*}
where $t_1^C < t_2^C < ... < t_{N_t}^C$ forms a grid on the $t$-domain with $N_t$ grid points, and $\phi_{i,1}^C < \phi_{i,2}^C < ... < \phi_{i,N_{\phi_i}}^C$ forms a grid on the $\phi_i$-domain with $N_{\phi_i}$ points. The collection of all points $(t^C, \phib^C)$ forms the collocation grid for training features. The choice of such points must satisfy $[T_0, T_1] \subseteq [t_1^C, t_{N_t}^C]$ and $[\phi_i^L, \phi_i^U] \subseteq [\phi_{i,1}^C, \phi_{i, N_{\phi_i}}^C]$ for all $i = 1, 2, ..., D+Q$ so that the grid is inclusive of all time points and parameter values of interest. To ensure good approximation, the collocation grid for training features should also be reasonably dense (i.e. $|t_{j+1}^C - t_{j}^C|$ and $|\phi_{i, j+1} - \phi_{i, j}|$ should be reasonably small) - achieved by taking $N_t$ and $N_{\phi_i}$ reasonably large - and somewhat evenly spaced. The number of collocation points $(t^C, \phib^C)$ arising from the above gridding exercise is therefore $N_t \cdot \prod_{i=1}^{D+Q} N_{\phi_i}$.\\

To obtain training targets for Method I, numerical solutions $\xb(t^C, \phib^C)$ are evaluated at each point $(t^C, \phib^C)$ of the training features' collocation grid, after which scaling is carried out to ensure that the targets are evenly trained. Define $F_d$ as some appropriate, invertible transformation (explained later) and set training targets as $F_d(x_d(t^C, \phib^C))$. Hence all sets of values $\{(t^C, \phib^C), F_d(x_d(t^C, \phib^C))\}$ constitute the training example for the individual network $\xtil_d(t, \phib)$ that approximates $x_d(t, \phib)$, and doing the same for all $d$ yields the full ANN that approximates $\xtilb(t, \phib)$. Training targets for Method II are constructed similarly, but with $x_d^\dagger$ replacing $x_d$ for all $d = 1, ..., D$. \\

One common choice for $F_d$ is the identity function, but in our investigations we found that the ANN performs better function approximations on a multivariable $\xtilb$ vector if scaling is performed on each component. To do this, $F_d$ evaluations are chosen to come from the $z$-score standardization function, i.e.
\begin{align*}
	F_d(x) := \frac{x - m_{x_d}}{s_{x_d}},
	\numberthis
	\label{eq:standardization}
\end{align*}
where $m_{x_d}$ and $s_{x_d}$ are the mean and standard deviation, respectively, of the collection of numerical solutions $x_d(t^C, \phib^C)$ arising from all points $(t^C, \phib^C)$ in the collocation grid; all targets $\{x_d\}_{d = 1, ..., D}$ are scaled using this method. Thus the inverse $F_d^{-1}$ can be written as
\begin{align*}
	F_d^{-1}(F) = s_{x_d} \cdot F + m_{x_d}.
	\numberthis
	\label{eq:destandardization}
\end{align*}
In our setting, it is important for $F_d$ to be invertible since reverting the ANN outputs $\Ftil_d$ to their original scale corresponding to the ODE model $\xtil_d$ will be crucial for the calculation of likelihoods detailed in Section~\ref{subs:introlaplace}. This target scaling is similarly applied to $\{x_d^\dagger\}_{d = 1, ..., D^\dagger}$ in Method II, denoted by $F_d^\dagger$ (with inverse ${F_d^\dagger}^{-1}$).

\subsection{Artificial Neural Networks: Architecture}
\label{subs:introdnn}

It is known that multilayer feed-forward neural networks with continuous sigmoidal activation functions are universal approximators to any measurable function (with degrees of success affected by the depth and width of the network) \cite{hornik1989multilayer}. For this reason, the function space for $\xtilb$ we will consider in this paper is that resulting from multilayer feed-forward neural networks with non-linear sigmoidal activations. \\

Since we wish to model (i.e. approximate) each $\xtil_d, d = 1, ..., D,$ as a function of time $t$ and the parameters $\phib$, the ANN should take these as input. In light of the target scaling mentioned in Equations (\ref{eq:standardization} - \ref{eq:destandardization}) in Section~\ref{subs:introdnn}, it is natural to define the output $\Ftil_d(t, \phib)$ as an $F_d$-transformed version of $\xtil_d$; in other words $\xtil_d(t, \phib)$ can simply be obtained as $F_d^{-1}(\Ftil_d(t, \phib))$. Mathematically, the neural network for Method I can be represented by
\begin{figure}
	\centering
	\includegraphics[width = 15cm]{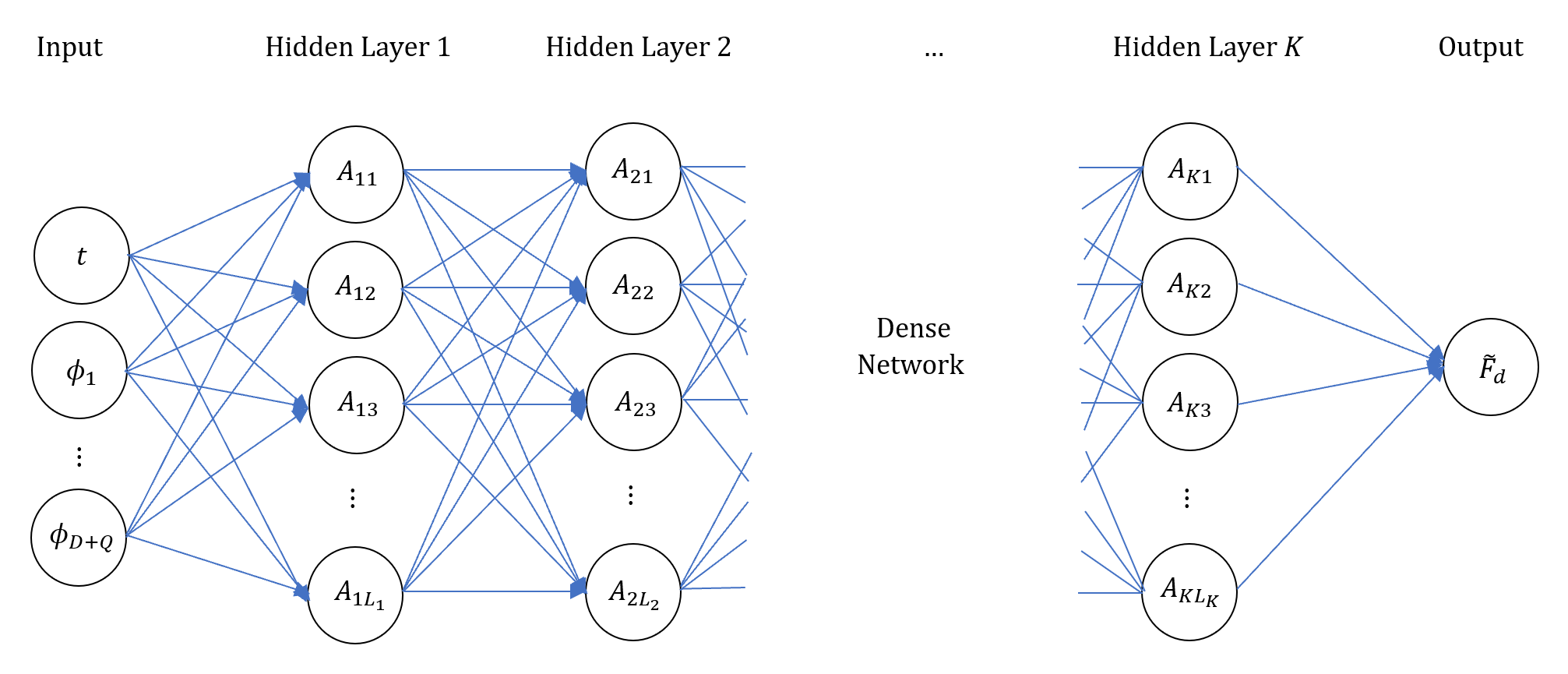}
	\caption{\label{fig:dnnarch} Architecture of the ANN for approximating $x_d(t, \phib), d= 1, ..., D$ (in Method I) or $x_d^\dagger(t, \phib), d= 1, ..., D^\dagger$ (in Method II). The input remains the same for both Methods I and II.}
\end{figure}
\begin{align*}
	A_{k,j} = \varphi{\Bigg(\sum_{i=1}^{L_{k-1}} w_{ij}^{(k)} A_{k-1, i} + b_j^{(k)}\Bigg)}, \hspace{0.5cm} & k = 1, 2, ..., K, \\
	\Ftil_d(t, \phib) = \sum_{i=1}^{L_K} w_{i1}^{(K+1)} A_{K,i} + b_1^{(K+1)}
	\numberthis
	\label{eq:ann}
\end{align*}
for $d = 1, 2, ..., D$ and a continuous sigmoidal function $\varphi$, where $w_{ij}^{(k)}$ is the weight connecting the $i$-th node of hidden layer $k-1$ to the $j$-th node of hidden layer $k$ and $b_j^{(k)}$ is the corresponding bias, for $k = 1, 2, ..., K+1$ (with the $0$-th layer referring to the input layer and $(K+1)$-th the output layer). For simplicity we may also rewrite the input nodes $(t, \phi_1, ... \phi_{D+Q})$ as $(A_{01}, A_{02}, ..., A_{0, D+Q+1})$. Equation (\ref{eq:ann}) can also be expressed vectorially:
\begin{align*}
	\mathbf{A}_k = \varphi
	\left(W^{(k)} \mathbf{A}_{k-1} + \mathbf{b}^{(k)}\right), \hspace{0.5cm} & k = 1, 2, ..., K, \\
	\Ftil_d(t, \phib) = W^{(K+1)}\mathbf{A}_K + \mathbf{b}^{(K+1)}, \hspace{0.5cm} & d = 1, 2, ..., D,
	\numberthis \label{eq:annvect}
\end{align*}
where $\varphi$ is taken component-wise. Finally, we concatenate the output $\Ftil_d$ of each individual ANN into one vector $\mathbf{\Ftil} := (\Ftil_1, \Ftil_2, ..., \Ftil_D)$ from which the loss function extracts the output values and backpropagates the loss into each individual ANN.\\

\begin{figure}
	\centering
	\includegraphics[width = 16cm]{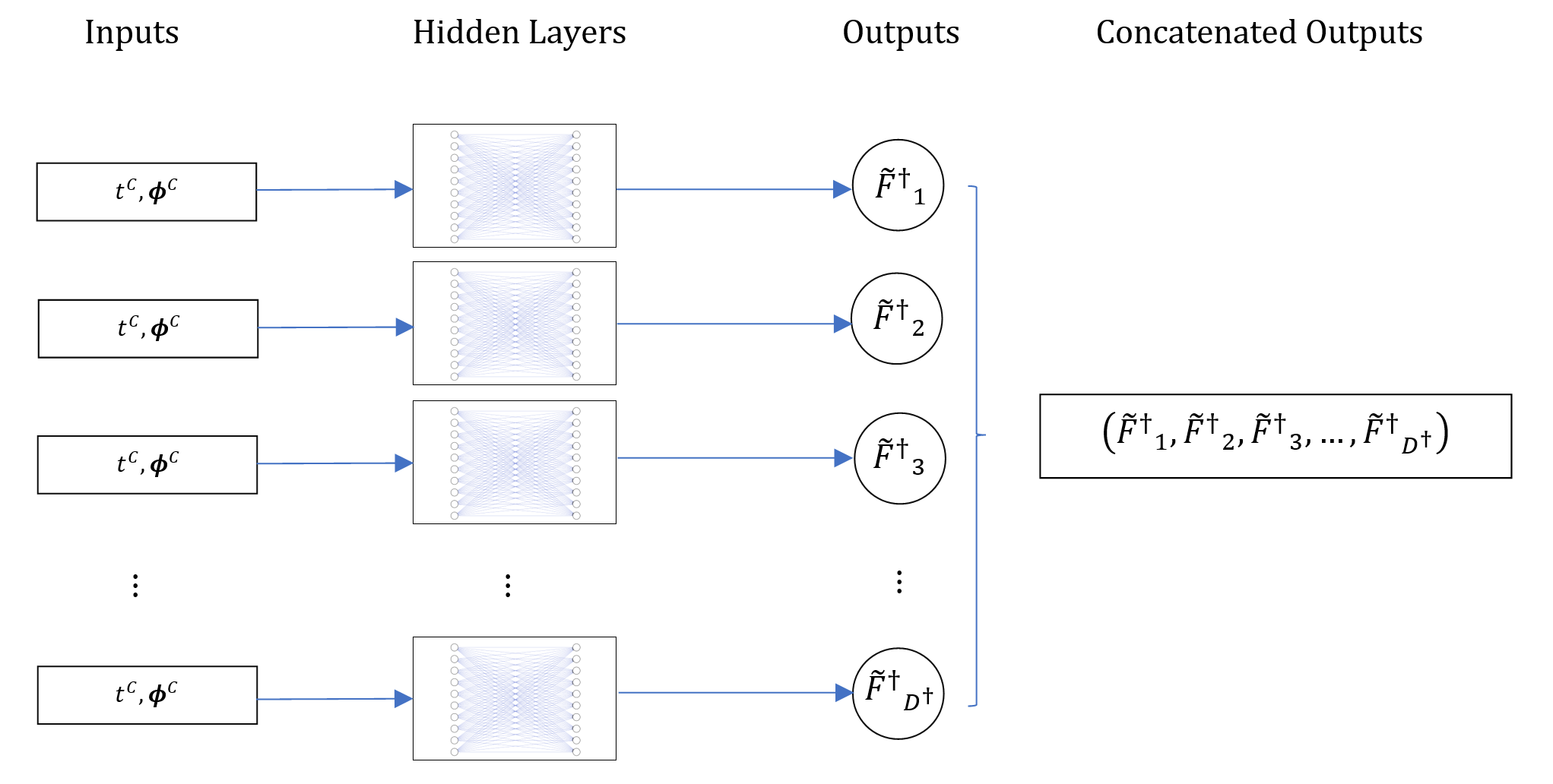}
	\caption{\label{fig:dnnarchconcat} The general architecture of concatenated ANNs used to approximate $\xtil_d^\dagger(t, \phib)$ for $d = 1, 2,..., D^\dagger$ used for Methods I and II.}
\end{figure}

Method II aims to also train the ANN to approximate first derivatives of the system states. To this end, we simply use more individual ANNs of the same architecture, each used to approximate each derivative component of $\xb^\dagger$ in Equation (\ref{eq:xdagger}). $\frac{\pa x_d}{\pa \phi_i}$. Since no modifications were made to the architecture, we may directly use $\xtil_d^\dagger$ and $\Ftil_d^\dagger$ in place of $\xtil_d$ and $\Ftil_d$ in Equations (\ref{eq:ann}) and (\ref{eq:annvect}), and extend them up to $d = D^\dagger = D(D+Q+1)$, i.e.
\begin{align*}
	\mathbf{A}_k = \varphi
	\left(W^{(k)} \mathbf{A}_{k-1} + \mathbf{b}^{(k)}\right), \hspace{0.5cm} & k = 1, 2, ..., K, \\
	\Ftil_d^\dagger(t, \phib) = W^{(K+1)}\mathbf{A}_K + \mathbf{b}^{(K+1)}, \hspace{0.5cm} & d = 1, 2, ..., D^\dagger.
	\numberthis \label{eq:annvect2}
\end{align*}
where $\varphi$ is again taken component-wise. The concatenation of outputs into a single vector $\mathbf{\Ftil}^\dagger := (\Ftil^\dagger_1, \Ftil^\dagger_2, ..., \Ftil^\dagger_{D^\dagger})$ in Figure~\ref{fig:dnnarchconcat} is also carried out to facilitate backpropagation; details are discussed in Section~\ref{subs:lossfn}. The general architecture of the ANNs used in Methods I and II are illustrated in Figure~\ref{fig:dnnarch}. Note that the inputs to the ANN architecture in both methods remain the same. Figure~\ref{fig:dnnarchconcat} gives the general architecture for the ANNs developed in Equation (\ref{eq:annvect2}) for each collocation point $(t^C, \phib^C)$.\\

Using these network architectures, $\xtilb(t, \phib)$ and $\xtilb^\dagger(t, \phib)$ specify analytically tractable functions of $t$ and $\phib$, which we can use to replace $\xb(t, \phib)$ in the likelihood (Equation (\ref{eq:generallikelihood})) as an approximation. For this approximation to be accurate we will need to train the network to find optimal weights $w_{ij}^{(k)}$ and biases $b_j^{(k)}$ based on suitable loss functions. \\

\subsection{Artificial Neural Network: Loss Function}
\label{subs:lossfn}

During training of the ANNs, loss functions are minimized with respect to the network weights $w_{ij}^{(k)}$ and biases $b_j^{(k)}$ using an optimization algorithm such as gradient descent algorithms or their adapted versions. Naturally, different loss functions lead to different optimal network weights and biases, as well as different outputs, thereby modifying the ANN's modelling properties. A common loss function used is the mean squared error (MSE) between training targets and the corresponding network output: 
\begin{align*}
	J_{MSE} = \frac{1}{M} \sum_{t', \phib'} \|\mathbf{F}(t', \phib') - \mathbf{\Ftil}(t',\phib')\|^2,
	\label{eq:mseloss}
	\numberthis
\end{align*}
where $\|\cdot\|$ is the $L^2$-norm and $\sum_{t',\phib'}$ denotes the sum taken over all $t$ and $\phib$ in each batch of training examples of size $M$. Equation~(\ref{eq:mseloss}) constitutes only a part of our loss function; the remaining parts attempt to improve the ANN model's adherence to the ODE system dynamics. The specific choices made are motivated subsequently. \\

Certain ODE models in the literature, for example, the SIR model by Kermack and McKendrick \cite{kermack1927contribution} for modelling infectious diseases satisfy a natural constraint among the components of $\xb(t, \phib)$. Let these natural constraints to the ODE system we wish to model be generally represented as 
\begin{align*}
	\mathbf{B}(\xb(t, \phib), \phib) = \mathbf{0}.
	\numberthis
	\label{eq:constraints}
\end{align*}
Corresponding to the extension in Equation (\ref{eq:dynsysextended}), we may also define an extended set of constraints $\mathbf{B}^\dagger(\xb^\dagger(t, \phib), \phib) = \mathbf{0}$, where
\begin{align*}
	\mathbf{B}^\dagger(\xb^\dagger(t, \phib), \phib) := 
	\begin{pmatrix}
		\mathbf{B}(\xb(t, \phib), \phib) \\
		\nabla_{\phi_1} \ \mathbf{B}(\xb(t, \phib), \phib) \\
		\nabla_{\phi_2} \ \mathbf{B}(\xb(t, \phib), \phib) \\
		\vdots \\
		\nabla_{\phi_{D+Q}} \ \mathbf{B}(\xb(t, \phib), \phib) \\
	\end{pmatrix},
\end{align*}
which are obtained by differentiating the original constraint equation (\ref{eq:constraints}) by $\phi_i$, for $i = 1, 2, ..., D+Q$. \\

For Method I, define the loss function $J_1$ using the $L^2$-norm $\|\cdot\|$:
\begin{align*}
	J_1 & = \frac{1}{M} \sum_{t', \phib'} \Bigg\{
	\left\| \mathbf{F}(t', \phib') - \tilde{\mathbf{F}}(t', \phib') \right\|^2 + 
	\left\| \lambdab_{11} \cdot \left(\frac{\pa \xtilb}{\pa t}(t', \phib') - \fb(\xtilb(t', \phib'), \phib')\right) \right\|^2 \\
	& \qquad \qquad \qquad \qquad
	+ \left\| \lambdab_{12} \cdot \mathbf{B}\big( \xtilb(t', \phib'), \phib' \big) \right\|^2 
	\Bigg\}
	\numberthis
	\label{eq:lossmethod1L2}
\end{align*}
and similarly, define the loss function $J_2$ for Method II by replacing $\xb$ with $\xb^\dagger$:
\begin{align*}
	J_2 & = \frac{1}{M} \sum_{t', \phib'} \Bigg\{
	\left\| \mathbf{F}^\dagger(t', \phib') - \tilde{\mathbf{F}}^\dagger(t', \phib') \right\|^2 + 
	\left\| \lambdab_{21} \cdot \left(\frac{\pa \xtilb^\dagger}{\pa t}(t', \phib') - \fb^\dagger(\xtilb^\dagger(t', \phib'), \phib')\right) \right\|^2 \\
	& \qquad \qquad \qquad \qquad
	+ \left\| \lambdab_{22} \cdot \mathbf{B}^\dagger\big( \xtilb^\dagger(t', \phib'), \phib' \big) \right\|^2 
	\Bigg\}
	\numberthis
	\label{eq:lossmethod2L2}
\end{align*}
where $\mathbf{F} = (F_1, F_2, ..., F_D)$ and $\mathbf{F}^\dagger = (F_1^\dagger, F_2^\dagger, ..., F_{D^\dagger}^\dagger)$ are the standardized targets based on the corresponding ANN output vectors, and $\lambdab_{11}, \lambdab_{12}, \lambdab_{21}, \lambdab_{22} \ge \mathbf{0}$ are vectors of regularization coefficients. \\

In both $J_1$ and $J_2$, the first term represents a mean squared error which penalizes the deviation of the ANN's approximation from the numerical solution (i.e. the training target). When $\lambdab_{11}, \lambdab_{21}> \mathbf{0}$, the second term attempts to explicitly force the system states and its partial derivatives to adhere to the ODE dynamics, similar to the derivative penalties applied in \cite{raissi2019physics} (except our ANNs are supervised by numerical solutions). The last term involving $\lambdab_{12}, \lambdab_{22}> \mathbf{0}$ represents the ANN approximations' adherence to the system's constraints. Note that, when $\lambdab_{11}, \lambdab_{12}, \lambdab_{21}$ and $\lambdab_{22}$ are identically $\mathbf{0}$, the defined loss functions reduce to a simple MSE between the ANN's outputs and training targets in Equation~(\ref{eq:mseloss}). Since $\xtilb$ and $\xtilb^\dagger$ are functions of $w_{ij}^{(k)}$ and $b_j^{(k)}$ during the training stage, the loss is backpropagated through each ANN and minimized with respect to their weights and biases. The concatenation illustrated in Figure~\ref{fig:dnnarchconcat} facilitates the backpropagation in an algorithmic way. \\

With the batch size, loss functions, and $\lambdab_{11}, \lambdab_{12}, \lambdab_{21}$ and $\lambdab_{22}$ specified, we can use an optimization algorithm to minimize the loss function with respect to each ANN's weights $w_{ij}^{(k)}$ and biases $b_j^{(k)}$, over a large number of epochs. The resulting weights and biases give rise to the outputs $\xtilb(t, \phib)$ (resp. $\xtilb^\dagger(t, \phib)$) which approximate the solution $\xb(t, \phib)$ (resp. $\xtilb^\dagger(t, \phib)$), and is used in the likelihood and posterior expressions.

\subsection{Laplace's Method}
\label{subs:introlaplace}

In general, Laplace's Method \cite{azevedo1994laplace} is an integral approximation technique. Its application to Bayesian statistics allows us to approximate the posterior distribution as a Gaussian centered around the \textit{maximum a posteriori} (MAP) estimate - the parameter value at which the posterior density is highest. \\

As shown in Equation (\ref{eq:bayesrule}), the exact posterior density $p(\phib|\yb)$ depends on the prior $p(\phib)$ which we may specify, and the exact likelihood $p(\yb|\phib) \equiv p(\yb|\xb(\tb, \phib), \phib)$. Note that the latter is intractable for ODE systems due to the intractability of $\xb(t, \phib)$. Thus, numerical solutions may be used as a substitute, but they possess the drawbacks discussed in Section \ref{subs:challenge}. We, therefore, propose using the ANN approximation to the solution, that is, we replace the likelihood term $p(\yb|\xb(\tb, \phib), \phib)$ with $p(\yb|\xtilb(\tb, \phib), \phib)$, thus giving it tractability for differentation. Despite this, $\xtilb(t, \phib)$ (and hence $p(\yb|\xtilb(\tb, \phib), \phib)$) is still a rather convoluted function of $t$ and $\phib$ - evaluating the full posterior would still prove to be difficult. To facilitate the derivation of the posterior distribution, we introduce another approximation using Laplace's method. We start by defining $g(\phib)$ such that $e^{g(\phib)} = p(\phib) p(\yb|\xtilb(\tb, \phib), \phib)$, and then approximate Equation (\ref{eq:bayesrule}) using $g(\phib)$:

\begin{align*}
	p(\phib|\yb) & = \frac{p(\phib) \ p(\yb|\xb(\tb, \phib), \phib)}{\int p(\phib) \ p(\yb| \xb(\tb, \phib), \phib) \ d\phib} \\
	& \approx \frac{p(\phib) \ p(\yb|\xtilb(\tb, \phib), \phib)}{\int p(\phib) \ p(\yb|\xtilb(\tb, \phib), \phib) \ d\phib} \tag*{(ANN approximation)} \\ 
	& = \frac{\exp\left(g(\phib)\right)}{\int \exp(g\left(\phib\right)) \ d\phib} \\
	& \approx \frac
	{\exp\left\{ g(\hat{\phib}) + \frac{1}{2}(\phib - \hat{\phib})^T \ \nabla_{\phib}^2 g(\hat{\phib}) \ (\phib - \hat{\phib}) \right\}}
	{\int \exp \left\{g(\hat{\phib}) + \frac{1}{2}(\phib - \hat{\phib})^T \ \nabla_{\phib}^2 g(\hat{\phib}) \ (\phib - \hat{\phib})\right\} \ d\phib } \tag*{(Taylor's expansion)} \\
	& = \frac
	{\exp \left\{\frac{1}{2}(\phib - \hat{\phib})^T \ \nabla_{\phib}^2 g(\hat{\phib}) \ (\phib - \hat{\phib})\right\}}
	{\int \exp \left\{\frac{1}{2}(\phib - \hat{\phib})^T \ \nabla_{\phib}^2 g(\hat{\phib}) \ (\phib - \hat{\phib})\right\} \ d\phib} \\
	& = \frac{1}{\sqrt{(2\pi)^{D+Q} \big|\det\big(-\nabla_{\phib}^{-2}g(\hat{\phib})\big)\big|}} \ 
	e^{-\frac{1}{2}(\phib-\hat{\phib})^T \big(-\nabla_{\phib}^{-2} g(\hat{\phib})\big)^{-1} (\phib - \hat{\phib})} \\
	& =: h(\phib|\yb)
\end{align*}

where $\hat{\phib}$ satisfies $\nabla_{\phib} g(\hat{\phib}) = 0$ and $h(\phib|\yb)$, a Gaussian density centered around $\hat{\phib}$ with variance $-\nabla_{\phib}^{-2} g(\hat{\phib})$, denotes the approximate posterior we seek. Hence 
\begin{align*}
	h(\phib|\yb) \sim N\left( \hat{\phib}, -\nabla_{\phib}^{-2} g(\hat{\phib}) \right).
	\numberthis
	\label{eq:posterior}
\end{align*}

The remaining work is to find $\hat{\phib}$ by optimizing $g({\phib})$ with respect to $\phib$. We can carry this out using an optimization algorithm; we opt to use a quasi-Newton algorithm called the Broyden–Fletcher–Goldfarb–Shanno (BFGS) algorithm \cite{fletcher2013practical}, which requires the use of the first derivatives of $g(\phib)$ with respect to $\phib$, $\nabla_{\phib} g(\phib) = \nabla_{\phib} \big( \ln p(\phib) + \ln p(\yb|\xtilb(\mathbf{t}, \phib), \phib) \big)$ - this is available as long as $p(\yb|\xtilb(\mathbf{t}, \phib), \phib)$ is a tractable and differentiable expression of $\xtilb(t, \phib)$ and $\phib$. This elucidates the need for accurate and smooth predictions of the first derivatives $\nabla_{\phib} \xtilb(t, \phib)$, and signifies the inclusion of the second term (with $\lambdab_{11}, \lambdab_{21} > \mathbf{0}$) in Equations (\ref{eq:lossmethod1L2}) and (\ref{eq:lossmethod2L2}). Furthermore, smoothed prediction surfaces of the said first derivatives lead to more reliable estimates for the Hessian $\nabla_{\phib}^2 \xtilb(t, \phib)$, which is crucial in the calculation of the variance $-\hessinv_{\phib} g(\hat{\phib})$ of the proposal distribution $h(\phib|\yb)$.\\

\begin{figure}
	\centering
	\includegraphics[width = 10cm]{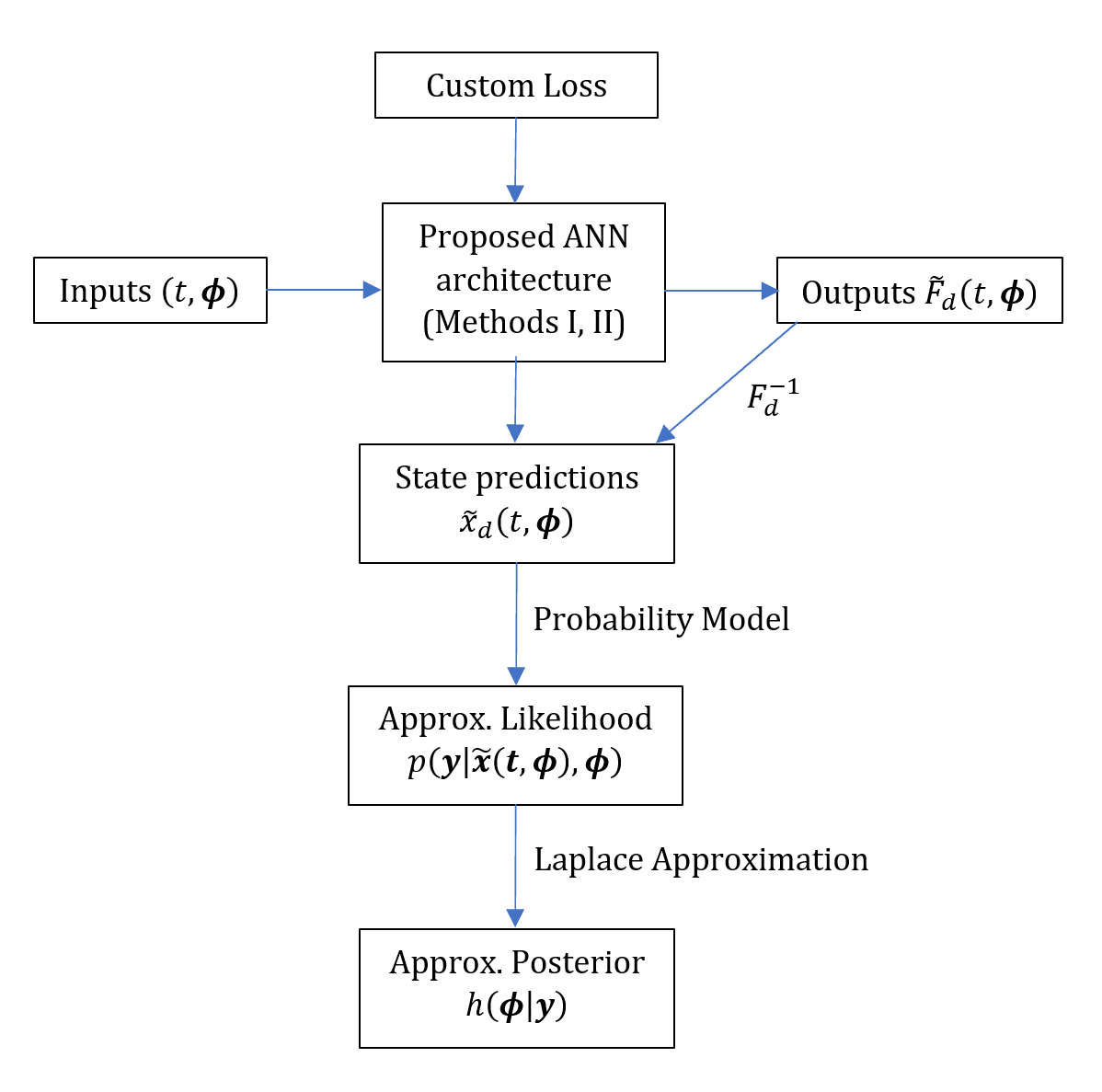}
	\caption{\label{fig:methodsummary} Flowchart of proposed methodology.}
\end{figure}

Figure~\ref{fig:methodsummary} shows a flowchart that summarizes the general algorithm for our methods which can be applied to ODE models arising from various dynamical systems. The next section illustrates the general methodology development for a special ODE model which is the SIR epidemic model.

\section{Application: The SIR Model}
\label{section:methods}

This section details the application of the proposed general methodology on a specific dynamical system: the Susceptible-Infected-Removed (SIR) model for infectious epidemics, first proposed by Kermack and McKendrick \cite{kermack1927contribution}. We first provide a brief introduction to the model's states and parameters, and then simulate datasets based on which we will draw posterior inference upon. The specific details of the ANN and Laplace Approximations are discussed subsequently. Before showing the results, we also describe the Metropolis Hastings (MH) algorithm which will act as a benchmark for our methods.

\subsection{The SIR Model}
\label{subs:sirmodel}

The SIR model is a compartmental model in epidemiology consisting of three compartments, or states, denoted by Susceptible ($S$), Infectious ($I$) and Removed ($R$). An individual in the population can move from a susceptible state, $S$, to an infectious state, $I$, upon infection, and after a certain infectious period, move to the removed state, $R$; it is assumed that those in the removed state do not return to the susceptible state. We introduce the following notation:
\begin{align*}
	P  & = \text{population size}, \\
	S(t) & = \text{number of susceptible individuals at time } t, \\
	I(t) & = \text{number of infectious individuals at time } t, \\
	R(t) & = \text{number of removed individuals at time } t, \\
	\beta & = \text{transmissibility of the disease} \\ & = \text{(average number of contacts per individual per unit time) } \times \\ & \qquad \text{(probability of a successful transmission)}, \\
	\gamma & = \text{average transition rate of an individual from infectious to removed}.
\end{align*}
Given these parameters, the SIR ODE model is then specified by the following system of ODEs:
\begin{align*}
	& \frac{dS(t)}{dt} = -e^{c_\beta} \frac{S(t) I(t)}{P}, \numberthis \label{eq:sir:S} \\[0.2cm]
	& \frac{dI(t)}{dt} = e^{c_\beta} \frac{S(t) I(t)}{P} - e^{c_\gamma} I(t), \numberthis \label{eq:sir:I} \\[0.2cm]
	& \frac{dR(t)}{dt} = e^{c_\gamma} I(t),  \numberthis \label{eq:sir:R} \\[0.2cm]
	& S(t) + I(t) + R(t) = P \ \ \ \forall t \in [t_0, t_N], \numberthis \label{eq:sir:sum} \\[0.2cm]
	& S(0) = P - e^{c_{I_0}}, \ I(0) = e^{c_{I_0}}, \ R(0) = 0. \numberthis \label{eq:sir:init}
\end{align*}
where we have parameterized $c_{I_0} := \ln I(0)$, $c_\gamma := \ln \gamma$ and $c_\beta := \ln \beta$ to ensure that we infer positive values for $\beta, \gamma$ and $I(0)$. Here we assume $\beta$ and $\gamma$ are constant over all $t \in [T_0, T_1]$, as in the original SIR model by Kermack and McKendrick \cite{kermack1927contribution}. Thus $\phib = \cb := (c_{I_0}, c_\gamma, c_\beta)$ based on notation introduced in Section~\ref{section:prelim}. Note that $S(0) = P - e^{c_{I_0}}$ is not included in $\phib$ since it is simply a function of $c_{I_0}$ and $R(0)$ is fixed at 0.\\

Further differentiating equations (\ref{eq:sir:S}), (\ref{eq:sir:I}) and (\ref{eq:sir:R}) with respect to $c_{I_0}, c_\gamma$ and $c_\beta$ extends the ODE system for our methodology's needs. Abbreviating $S(t, \cb), I(t, \cb)$ and $R(t, \cb)$ as $S, I$ and $R$ respectively, we derive the following equations.
\begin{align*}
	\deldelt \frac{\pa S}{\pa c_{I_0}} 
	& = -\frac{1}{P} e^{c_\beta} \left( I \frac{\pa S}{\pa c_{I_0}} + S \frac{\pa I}{\pa c_{I_0}} \right) 
	\numberthis \label{eq:sir:dsdci0} \\
	\deldelt \frac{\pa S}{\pa c_{\gamma}}
	& = -\frac{1}{P} e^{c_\beta} \left( I \frac{\pa S}{\pa c_{\gamma}} + S \frac{\pa I}{\pa c_{\gamma}} \right) 
	\numberthis \label{eq:sir:dsdcgamma} \\
	\deldelt \frac{\pa S}{\pa c_{\beta}}
	& = -\frac{1}{P} e^{c_\beta} \left(SI + I \frac{\pa S}{\pa c_{\beta}} + S \frac{\pa I}{\pa c_{\beta}} \right) 
	\numberthis \label{eq:sir:dsdcbeta} \\
	\deldelt \frac{\pa I}{\pa c_{I_0}}
	& = \frac{1}{P} e^{c_\beta} \left( I \frac{\pa S}{\pa c_{I_0}} + S \frac{\pa I}{\pa c_{I_0}} \right) - e^{c_\gamma} \frac{\pa I}{\pa c_{I_0}}
	\numberthis \label{eq:sir:didci0} \\
	\deldelt \frac{\pa I}{\pa c_{\gamma}}
	& = \frac{1}{P} e^{c_\beta} \left( I \frac{\pa S}{\pa c_{\gamma}} + S \frac{\pa I}{\pa c_{\gamma}} \right) - e^{c_\gamma} \left(I+\frac{\pa I}{\pa c_{I_0}}\right)
	\numberthis \label{eq:sir:didcgamma} \\
	\deldelt \frac{\pa I}{\pa c_{\beta}}
	& = \frac{1}{P} e^{c_\beta} \left(SI + I \frac{\pa S}{\pa c_{\beta}} + S \frac{\pa I}{\pa c_{\beta}} \right) - e^{c_\gamma}\frac{\pa I}{\pa c_\beta}
	\numberthis \label{eq:sir:didcbeta} \\
	\deldelt \frac{\pa R}{\pa c_{I_0}}
	& = e^{c_\gamma} \frac{\pa I}{\pa c_{I_0}}
	\numberthis \label{eq:sir:drdci0} \\
	\deldelt \frac{\pa R}{\pa c_{\gamma}}
	& = e^{c_\gamma} \left(I+\frac{\pa I}{\pa c_{I_0}}\right)
	\numberthis \label{eq:sir:drdcgamma} \\
	\deldelt \frac{\pa R}{\pa c_{\beta}}
	& = e^{c_\gamma} \frac{\pa I}{\pa c_{\beta}}
	\numberthis \label{eq:sir:drdcbeta} 
\end{align*}
\begin{align*}
	\frac{\pa S}{\pa c_{I_0}} + \frac{\pa I}{\pa c_{I_0}} + \frac{\pa R}{\pa c_{I_0}} = 0 \numberthis \label{eq:sir:derivsumci0} \\
	\frac{\pa S}{\pa c_{\gamma}} + \frac{\pa I}{\pa c_{\gamma}} + \frac{\pa R}{\pa c_{\gamma}} = 0 \numberthis \label{eq:sir:derivsumcgamma} \\
	\frac{\pa S}{\pa c_{\beta}} + \frac{\pa I}{\pa c_{\beta}} + \frac{\pa R}{\pa c_{\beta}} = 0 \numberthis \label{eq:sir:derivsumcbeta}
\end{align*}

Applying Method I's notations to the SIR Model, $\xb = (S, I, R)$ and $\fb$ is clear from Equations (\ref{eq:sir:S}) to (\ref{eq:sir:R}). Equation (\ref{eq:sir:sum}) also gives the constraint $B(\xb(t, \cb), \cb) = S(t, \cb) + I(t, \cb) + R(t, \cb) - P = 0$. For Method II, $\xb^\dagger$ is defined as $\Big(S, I, R, \frac{\pa S}{\pa c_{I_0}}, \frac{\pa I}{\pa c_{I_0}}, \frac{\pa R}{\pa c_{I_0}}, \frac{\pa S}{\pa c_{\gamma}}, \frac{\pa I}{\pa c_{\gamma}}, \frac{\pa R}{\pa c_{\gamma}}, \frac{\pa S}{\pa c_{\beta}}, \\ \frac{\pa I}{\pa c_{\beta}}, \frac{\pa R}{\pa c_{\beta}} \Big)$, and $\fb^\dagger$ is evident from Equations (\ref{eq:sir:S})-(\ref{eq:sir:R}) and (\ref{eq:sir:dsdci0})-(\ref{eq:sir:drdcbeta}). $\mathbf{B}^\dagger(\xb^\dagger(t, \cb), \cb)$ can similarly be obtained from Equations (\ref{eq:sir:derivsumci0}) to (\ref{eq:sir:derivsumcbeta}).
With these, we are ready to formulate our ANN architecture, loss functions and training algorithms for the SIR Model.

\subsection{Simulated Datasets}
\label{subs:sirsimdata}

In order to assess the approximations to the true posterior over a number of experiments, we simulate 200 different sets of data $\{\yb_i\}_{i = 1, 2, ..., 200}$ where each $\yb_i := (y_i(t_1), ..., y_i(t_N))$ is generated from a fixed set of parameters $\cb^* = (c_{I_0}^*, c_\gamma^*, c_\beta^*) = (\ln 7, \ln 1/7, -0.25)$ (henceforth referred to as the "true" value of the ODE's parameters) and $P = 10,000$. As in real practice, only some of the $S$-$I$-$R$ compartments will be observed. In this simulated experiment, we choose $y_i(t_n)$ to represent the number of new removed (state $R$) individuals within the time interval $(t_{n-1}, t_n]$; further setting $t_n = n \text{ days}$ means the $y_i(t_n)$'s represent daily numbers of newly removed individuals. Since $\frac{dR}{dt} = e^{c_\gamma}I(t, \cb)$ is the corresponding rate, we may use it as the expected value of $y_i(t_j)$. Hence, for all $i = 1, 2, ..., 200$, $y_i(t_n)$ is randomly generated from a Poisson distribution with mean $e^{c_\gamma^*} I(t_n, \cb^*)$, where $I(t, \cb^*)$ is the numerical solution of the state $I$ at time $t$ under the parameter setting $\cb^*$.

\subsection{ANN: Training Examples and Architecture}
\label{subs:dnnarch2}

For training features of both Methods I and II, we chose a grid of equally spaced collocation points based on the following:
\begin{align*}
	& t_1^C = 1, \ && t_{N_t}^C = 50, \ && N_t = 50; \\
	& c_{I_0, 1}^C = \ln 6, \ && c_{I_0, N_{c_{I_0}}}^C = \ln 8, \ && N_{c_{I_0}} = 16; \\
	& c_{\gamma, 1}^C = \ln 1/8, \ && c_{\gamma, N_{c_\gamma}}^C = \ln 1/6, \ && N_{c_\gamma} = 16; \\
	& c_{\beta, 1}^C = -0.3, \ && c_{\beta, N_{c_\beta}}^C = -0.2, \ && N_{c_\beta} = 16.
	\numberthis
	\label{eq:grid}
\end{align*}

Numerically solving the ODE for each parameter combination and then applying a target scaling transformation (i.e. $\mathbf{F}$ in Method I and $\mathbf{F}^\dagger$ in Method II) - the $z$-score standardization - gives rise to  $N_t \times N_{c_{I_0}} \times N_{c_\gamma} \times N_{c_\beta} = 50 \times 16^3 = 204,800$ examples. 20\% of these were randomly selected to form the test dataset, and subsequently a further 20\% of the remaining examples comprised the validation set (leaving $64\% \times 204,800 = 131,072$ examples for training). The choices for the values in (\ref{eq:grid}) are based on true values $\cb^* = (\ln 7, \ln 1/7, -0.25)$ of the parameters used to simulate datasets in Section~\ref{subs:sirsimdata}. Note that $N_t, N_{c_{I_0}}, N_{c_\gamma}$ and $N_{c_\beta}$ were chosen so that the grid is dense enough for the ANN to produce good approximations. \\

With reference to Figure \ref{fig:dnnarch} in Section~\ref{subs:introdnn}, the depths and widths chosen for each ANN were $K = 2$ and $L_1 = L_2 = 10$ respectively, along with $\tanh$ activations (i.e. $\varphi := \tanh$) for each hidden layer. The batch size chosen was $M = 400$ and the ANNs were trained for 2,500 epochs, using the loss functions $J_1$ and $J_2$ in Equations (\ref{eq:lossmethod1L2}) and (\ref{eq:lossmethod2L2}), applying the notation and methodology to the SIR Model as discussed in Section \ref{subs:sirmodel}. The optimal values for $\lambdab_{11}, \lambdab_{12}, \lambdab_{21}$ and $\lambdab_{22}$ were found by trial and error - the values that gave the least weighted absolute error in the approximation were chosen, although we found the performance did not vary largely within a small neighbourhood of these optimal values. The values chosen were $\lambdab_{11} = (0.5, 0.5, 0.5), \lambdab_{12} = 0.5, \lambdab_{21} = (0.5, 0.5,0.5,0.001,0.001,\\0.001,0.001,0.001,0.001,0.001,0.001,0.001)$ and $\lambdab_{22} = (0.5, 0.001, 0.001, 0.001)$.\\

All model building and training described above were done using the \texttt{tensorflow} and \texttt{keras} libraries in R. Through functions in these libraries we were able to obtain the derivatives of the outputs with respect to each of their inputs (such as $\frac{\pa}{\pa t}\Stil(t', \cb'), \frac{\pa}{\pa c_{I_0}}\Stil(t', \cb')$, etc.), as is required by the loss function defined above when $\lambda_1, \lambda_2 > 0$.

\subsection{Laplace Approximation}
\label{subs:2laplace}
\renewcommand\arraystretch{2}

Priors for $\cb$ were specified to be independently Gaussian:
\begin{align*}
	c_{I_0} \sim \text{N}\left(\mu_{c_{I_0}}, \sigma_{c_{I_0}}^2\right); && 
	c_{\gamma} \sim \text{N}\left(\mu_{c_{\gamma}}, \sigma_{c_{\gamma}}^2\right); &&
	c_{\beta} \sim \text{N}\left(\mu_{c_{\beta}}, \sigma_{c_{\beta}}^2\right),
	\numberthis \label{eq:prior}
\end{align*}
giving rise to the prior density
\begin{align*}
	p(\cb) & = 
	\prod_{\phi \ \in \ \{c_{I_0}, c_\gamma, c_\beta\}}
	\frac{1}{\sqrt{2\pi\sigma_\phi^2}} \  \exp\left\{-\frac{1}{2\sigma_\phi^2}\left(\phi - \mu_\phi\right)^2\right\} \\
	& \propto \exp
	\left\{
	-\frac{1}{2\sigma_{c_{I_0}}^2}\left(c_{I_0} - \mu_{c_{I_0}}\right)^2
	-\frac{1}{2\sigma_{c_{\gamma}}^2}\left(c_\gamma - \mu_{c_{\gamma}}\right)^2
	-\frac{1}{2\sigma_{c_{\beta}}^2}\left(c_\beta - \mu_{c_{\beta}}\right)^2
	\right\}
	\numberthis
	\label{eq:priordensity}
\end{align*}
and values for the means and variances were chosen to elicit vague \cite{gelman1995bayesian} priors:
\begin{align*}
	\mu_{c_{I_0}} = \mu_{c_{\gamma}} = \mu_{c_{\beta}} = 0; \ \ \  \sigma_{c_{I_0}}^2 = \sigma_{c_{\gamma}}^2 = \sigma_{c_{\beta}}^2 = 10^4
	\numberthis \label{eq:priorvalues}
\end{align*} 
Corresponding to the model used to generate the datasets $\{\yb_i\}_{i=1, 2, ..., 200}$, a Poisson model with mean $e^{c_\gamma}\Itil(t_n, \cb)$ was used for the likelihood of observing each $y_i(t_n), n = 1, 2, ..., N$. Noting the conditional independence of $y_i(t_n), n = 1, ..., N$ given $\cb$, the complete likelihood $\mathcal{L}(\cb; \yb_i)$ can be written as
\begin{align*}
	\mathcal{L}(\cb; \yb_i) &
	= \prod_{n = 1}^{N} p\left(y_i({t_n})\middle|\cb\right) 
	\approx \prod_{n = 1}^{N} \text{Poi}\left(y_i({t_n})\middle| e^{c_\gamma}\Itil(t_n,\cb)\right) 
	= \prod_{n = 1}^{N} 
	\frac
	{\left(e^{c_\gamma}\Itil(t_n,\cb)\right)^{y_i({t_n})} e^{-e^{c_\gamma}\Itil(t_n, \cb)}}
	{y_i(t_n)!}
	\numberthis
	\label{eq:likelihood}
\end{align*}
resulting in the log-likelihood
\begin{align*}
	\ell(\cb; \yb_i) & = \ln \mathcal{L}(\undc; \yb_i) \approx
	\sum_{n=1}^{N} y_i({t_n}) \ln \left(e^{c_\gamma}\Itil(t_n,\cb)\right) - \sum_{n=1}^{N} e^{c_\gamma}\Itil(t_n,\cb) - \sum_{n=1}^{N} \ln (y_i(t_n)!) \\
	& = \sum_{n=1}^{N} y_i({t_n}) \left({c_\gamma} + \ln \Itil(t_n, \cb)\right) - \sum_{n=1}^{N} e^{c_\gamma}\Itil(t_n,\cb) + \text{constant independent of } \cb,
	\numberthis
	\label{eq:loglikelihood}
\end{align*}
where $\Itil(t_n, \cb)$ is the approximation to the solution of state $I$ by the ANN specified in Section~\ref{subs:dnnarch2}. We highlight here that the \textit{same} set of ANNs is used for all 200 datasets without having to repeat training - this demonstrates the advantage of using function approximation methods. This feature thus has the added advantage of not needing reruns during the inference stage unlike in the case of Markov Chain Monte Carlo (MCMC) methods.\\

We can now formulate the Laplace approximation that would lead us to the appropriate proposal distribution/approximate posterior $h(\cb|\yb_i)$. As before, writing $e^{g(\cb)} = p(\cb)p(\yb_i|\cb)$ we find an expression for $g$ using Equations (\ref{eq:priordensity}) and (\ref{eq:loglikelihood}):
\begin{align*}
	g(\cb) & = \ln p(\cb) + \ln p(\yb_i|\cb) & =  -\frac{1}{2\sigma_{c_{I_0}}^2}\left(c_{I_0} - \mu_{c_{I_0}}\right)^2
	-\frac{1}{2\sigma_{c_{\gamma}}^2}\left(c_\gamma - \mu_{c_{\gamma}}\right)^2
	-\frac{1}{2\sigma_{c_{\beta}}^2}\left(c_\beta - \mu_{c_{\beta}}\right)^2 \\ & \ & \qquad
	+ \sum_{n=1}^{N} y_i({t_n}) \left({c_\gamma} + \ln \Itil(t_n, \cb)\right) - \sum_{n=1}^{N} e^{c_\gamma}\Itil(t_n, \cb) + \text{const.}
	\numberthis \label{eq:g}
\end{align*}
The main objective of the Laplace Approximation is to find the approximate \textit{maximum a posteriori} (MAP) $\hat{\cb}$ such that $g(\hat{\cb}) = 0$ - this was done using the BFGS optimization algorithm within the \texttt{optimx} package in R, with initial point on the boundary of the collocation grid $(\ln 8, \ln 1/6, -0.2)$. Finally, we arrive at the posterior Gaussian distribution $\cb \sim N(\hat{\cb}, -\nabla^{-2} g(\hat{\cb}))$ after observing the dataset $\yb_i$. The above procedure is repeated for each of the 200 experiments based on different sets of observations $\yb_i, i = 1, 2, ..., 200$, which can be completed very quickly. On a HP workstation with 16 GB RAM and 12 Intel Core 7.0 processors with processing speed of 2.60 GHz, each Laplace approximation task took, on average, 27.6 seconds for Method I and 1.77 seconds for Method II.

\subsection{Performance Benchmark}

The performance of our method will be compared with the Random-Walk Metropolis Hastings (MH) algorithm for sampling from the posterior distribution; see \cite{streftaris2004bayesian} for an example of such an application. For consistency, the same priors as in Equation (\ref{eq:prior}) were used along with the likelihood $p(\yb_i|\cb) = \prod_{n=1}^{N}\text{Poi}(y_i({t_n})|e^{c_\gamma}I(t_n,\cb))$ where $I(t,\cb)$ comes from the numerical solution for given a set of (proposed) parameters $\cb$, for all $i = 1, 2, ..., 200$. \\
	
Let $n_{iter}$ be the number of iterations of the algorithm, and $\alpha \in \mathbb{Z}^+$ be some positive integer that divides $n_{iter}$. The Random-Walk MH Algorithm for obtaining the posterior $h(\cb|\yb_i)$ arising from the dataset $\yb_i$ is as follows. 
\begin{itemize}
	\item Set an initial value of parameters $\cb^{(0)}$,
	\item Solve the system of ODEs for the SIR Model numerically (Equations (\ref{eq:sir:S}) - (\ref{eq:sir:init})) to obtain $I(t,\cb^{(0)})$,
	\item For $j = 1, 2, ..., n_{iter}$:
	\begin{itemize}
	\item Propose a new set of parameters $\cb'$ from a symmetrical distribution $\cb' \sim N(\cb^{(j-1)}, S^2)$ where $S^2 := diag(s_{c_{I_0}}^2, s_{c_\gamma}^2, s_{c_\beta}^2)$ controls the variance of this proposal,
	\item Solve the system of ODEs for the SIR Model to obtain $I(t, \cb')$,
	\item Calculate the probability of acceptance $p_{acc}$ as:
	\begin{align*}
		p_{acc} = \frac
		{p(\cb')\prod_{n=1}^{N}\text{Poi}\big(y_i({t_n})|e^{c_\gamma'}I(t_n,\cb')\big)}
		{p(\cb^{(j-1)})\prod_{n=1}^{N}\text{Poi}\big(y_i({t_n})|e^{c_\gamma^{(j-1)}}I(t_n, \cb^{(j-1)})\big)},
	\end{align*}
	\item Accept $\cb'$ as a sample from the posterior with probability $p_{acc}$ and set $\cb^{(j)} = \cb'$; otherwise reject $\cb'$ and set $\cb^{(j)} = \cb^{(j-1)}$ (i.e. we are employing a ``block'' MH algorithm where all parameters in $\cb$ are updated together, and not separately),
	\end{itemize}
	\item $\big\{\cb^{(\alpha)}, \cb^{(2\alpha)}, ..., \cb^{(n_{iter}-\alpha)}, \cb^{(n_{iter})} \big\}$ is taken as the posterior sample.
\end{itemize}
Sufficiently large $n_{iter}$ and $\alpha$ were chosen along with appropriate values of $s_{c_{I_0}}^2, s_{c_\gamma}^2, s_{c_\beta}^2$ to ensure good representation of the posterior: $n_{iter} = $. Further, $\cb^{(0)}$ was conveniently set to $\cb^*$ to avoid a long burn-in period. Applying a kernel density estimate to the posterior sample gives us the posterior density resulting from this Random-Walk MH algorithm. Note that the MH algorithm needs to be rerun for each $\yb_i, i = 1, 2, ..., 200$, for obtaining posterior inference on $\cb$. On the same computer specifications detailed at the end of Section~\ref{subs:2laplace}, each MH procedure took an average of 11.9 minutes even though the starting point $\cb^{(0)}$ was chosen very close to $\cb^*$.

\subsection{Results}
\label{subs:results}

In this subsection, we first show the accuracy of the ANN approximations $\xtilb$ and $\xtilb^\dagger$ from Methods I and II compared to the numerical solutions. Next, we examine the approximate posteriors $\{h(\cb|\yb_i)\}_{i = 1, 2, ..., 200}$ obtained from each method, compared to the benchmark from the MH algorithm. \\

Figure~\ref{fig:ANN1} shows the ANN approximations $\Stil(t, \cb^*), \Itil(t, \cb^*), \Rtil(t, \cb^*)$ as well as the derivatives obtained from the ANN approximations $\frac{\pa}{\pa c_{I_0}}\Stil(t, \cb^*), \frac{\pa}{\pa c_{I_0}}\Itil(t, \cb^*), ..., \frac{\pa}{\pa c_{\beta}}\Rtil(t, \cb^*)$ under Method I.  Figure~\ref{fig:ANN2} shows the ANN approximations $\Stil(t, \cb^*), \Itil(t, \cb^*), \Rtil(t, \cb^*), \tilde{\frac{\pa S}{\pa c_{I_0}}}(t, \cb^*), \\ \tilde{\frac{\pa I}{\pa c_{I_0}}}(t, \cb^*), ..., \tilde{\frac{\pa R}{\pa c_{\beta}}}(t, \cb^*)$ under Method II; note that unlike Method I, the approximations to the derivatives are directly available from the ANN outputs. The differences in approximation performances of $\Stil(t, \cb^*), \Itil(t, \cb^*), \Rtil(t, \cb^*)$ between Methods I and II are not obvious on the plots, however the approximations for the derivatives in Method II are evidently closer to the numerical solutions, as expected due to the explicit training and regularization; the plots for the second derivatives are not shown but follow a similar pattern to the first derivatives. To illustrate these differences quantitatively, Table~\ref{tab:sse} shows the percentage reduction, $p$, from the total sum of squares (TSS) by the ANNs in Methods I and II under the parameter setting $\cb^*$, where
\begin{align*}
	TSS(x_d, \cb^*) & := \sum_{t=1}^{50} \left(x_d(t, \cb^*) - \frac{1}{50}\sum_{t=1}^{50} x_d(t, \cb^*)\right)^2, \\
	SSE(x_d, \cb^*) & := \sum_{t=1}^{50} \left(x_d(t, \cb^*) - \xtil_d(t, \cb^*)\right)^2, \\
	p(c_d, \cb^*) & := \frac{TSS - SSE}{TSS} \times 100\%.
	\numberthis
	\label{eq:percred}
\end{align*}  
Here, TSS represents the error resulting from the simple mean estimate $\frac{1}{50}\sum_{t=1}^{50} x_d(t, \cb^*)$, and SSE represents the error between the ANN approximation $\xtil_d$ and the numerical solution $x_d$. $p$ therefore describes the extent of reduction in error provided by the ANN approximations relative to the simple mean estimate; a larger value of $p$ indicates better approximation by the ANN. \\

Notably, the percentage reductions from TSS achieved for all states and derivatives under Method II are larger than that under Method I; Method II achieves over 99\% reduction from TSS for all states, first derivatives and second derivatives, whereas Method I falls behind especially in approximating the second derivatives. This is in line with the design of the ANN in Method II - explicit training of the derivatives as well as more regularizations. The accuracy of such ANN approximations have an effect on the Laplace Approximation that follows, due to the usages of $\Itil(t, \cb)$ in the likelihood expression, $\nabla_{\cb} \Itil(t, \cb)$ in the BFGS algorithm, and $\nabla_{\cb}^2 \Itil(t,\cb)$ in quantifying the uncertainty of the MAP estimate. We now examine the performance of each method in estimating the posterior distribution. \\

{\renewcommand{\arraystretch}{0.5}

\begin{figure}
	\centering
	\begin{tabular}{ccc}
		\includegraphics[width = 5cm]{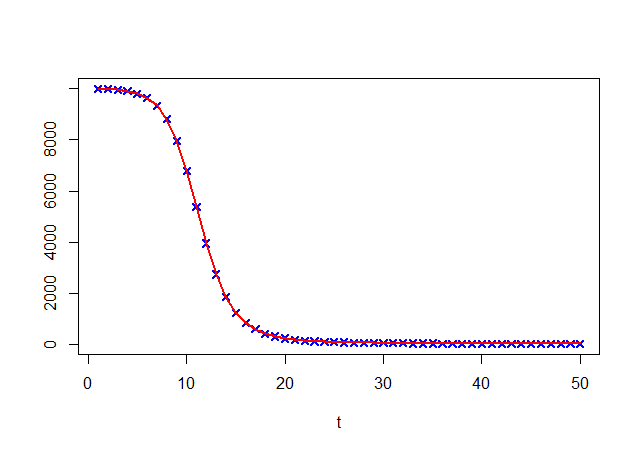} &
		\includegraphics[width = 5cm]{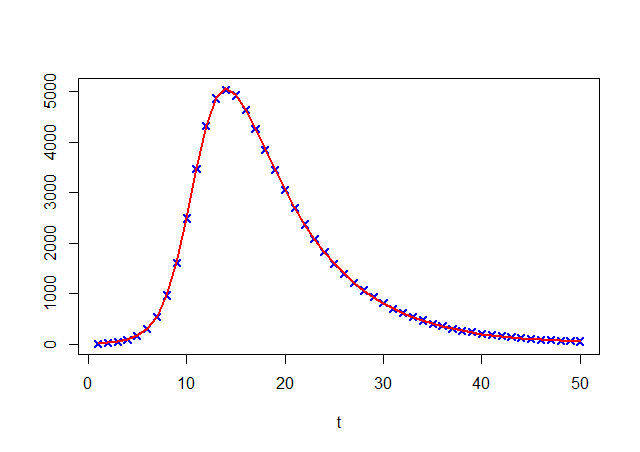} &
		\includegraphics[width = 5cm]{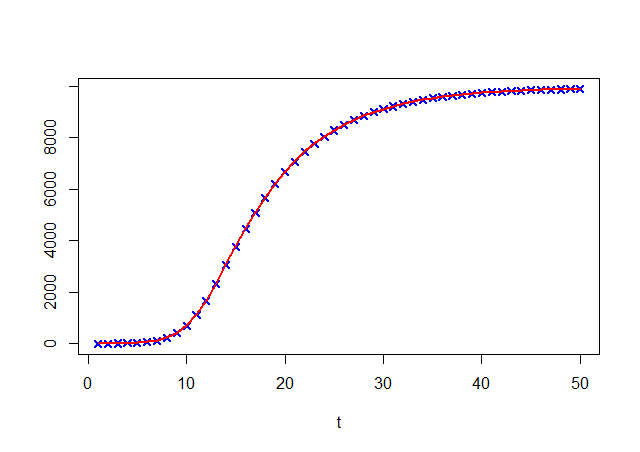} \\
		$\Stil(t, \cb^*)$ & $\Itil(t, \cb^*)$ & $\Rtil(t, \cb^*)$ \\[20pt]
		\includegraphics[width = 5cm]{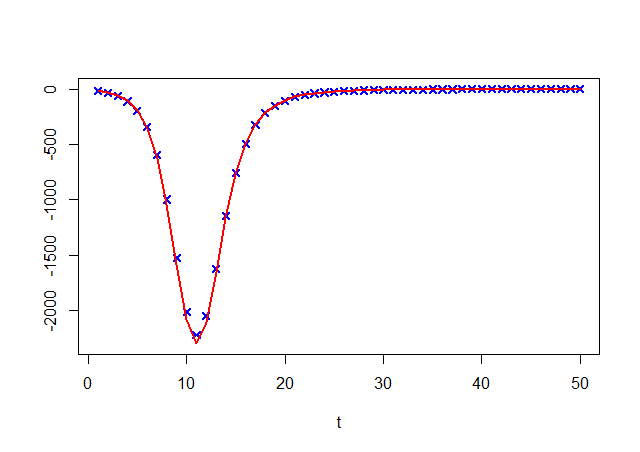} &
		\includegraphics[width = 5cm]{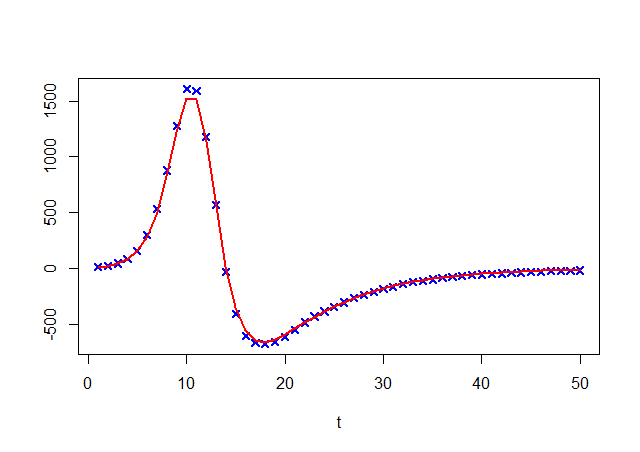} &
		\includegraphics[width = 5cm]{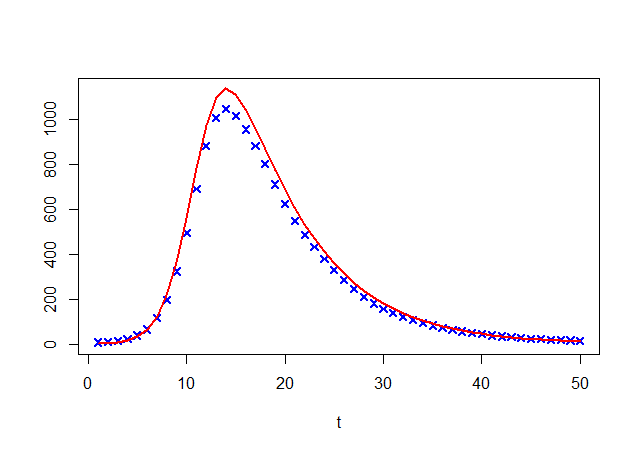} \\
		$\frac{\pa}{\pa c_{I_0}}\Stil(t, \cb^*)$ & $\frac{\pa}{\pa c_{I_0}}\Itil(t, \cb^*)$ & $\frac{\pa}{\pa c_{I_0}}\Rtil(t, \cb^*)$ \\[20pt]
		\includegraphics[width = 5cm]{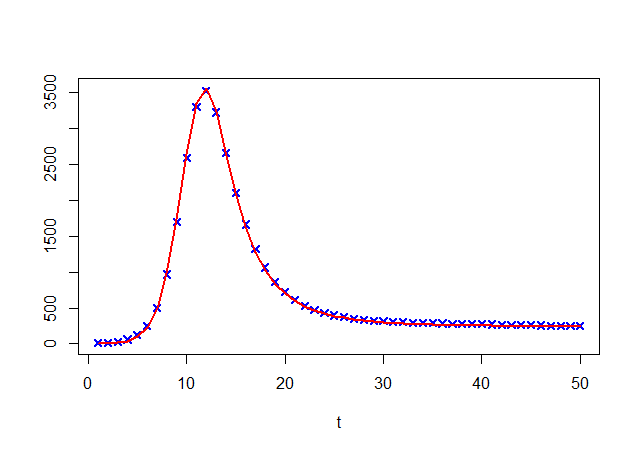} &
		\includegraphics[width = 5cm]{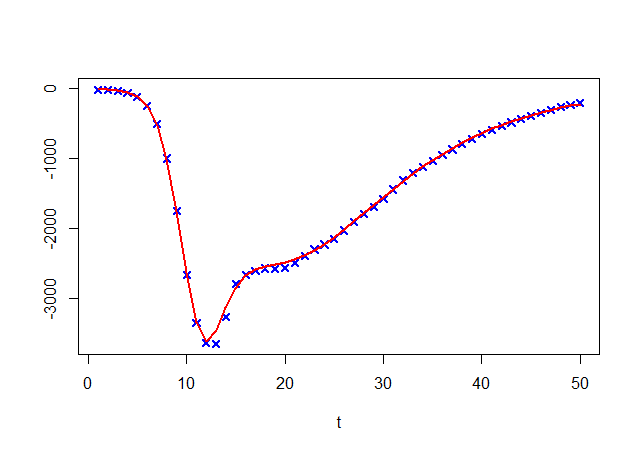} &
		\includegraphics[width = 5cm]{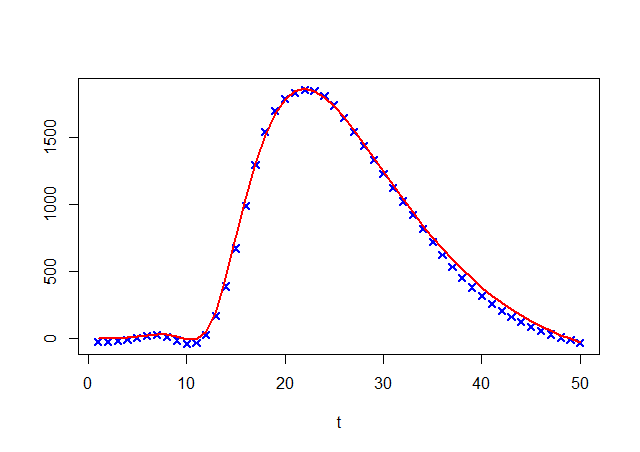} \\
		$\frac{\pa}{\pa c_{\gamma}}\Stil(t, \cb^*)$ & $\frac{\pa}{\pa c_{\gamma}}\Itil(t, \cb^*)$ & $\frac{\pa}{\pa c_{\gamma}}\Rtil(t, \cb^*)$ \\[20pt]
		\includegraphics[width = 5cm]{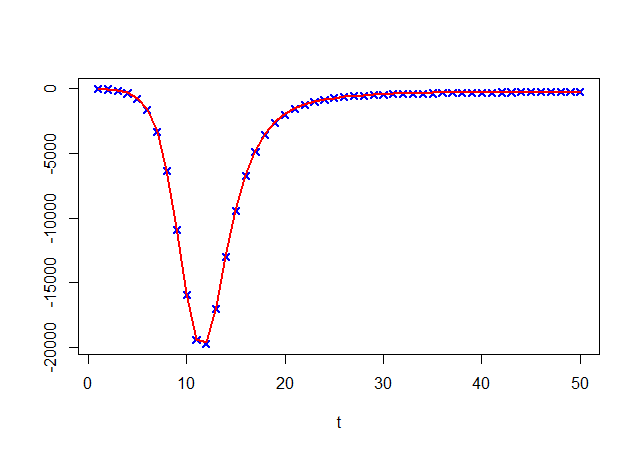} &
		\includegraphics[width = 5cm]{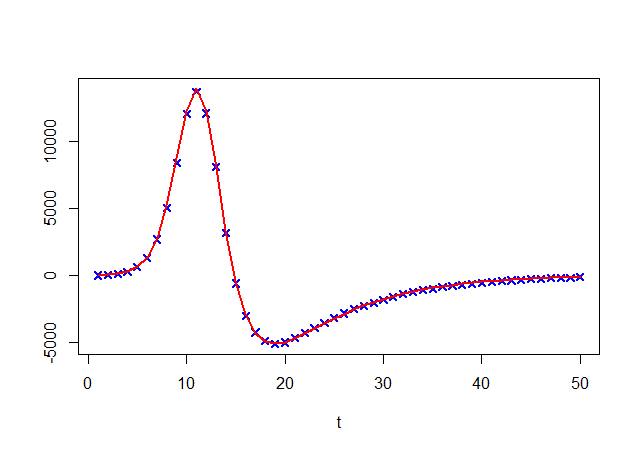} &
		\includegraphics[width = 5cm]{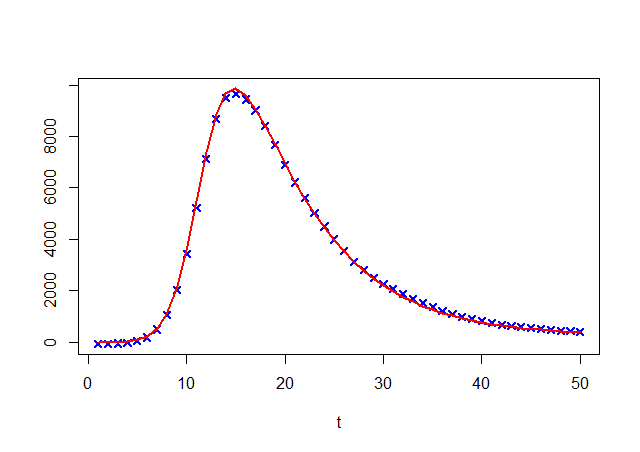} \\
		$\frac{\pa}{\pa c_{\beta}}\Stil(t, \cb^*)$ & $\frac{\pa}{\pa c_{\beta}}\Itil(t, \cb^*)$ & $\frac{\pa}{\pa c_{\beta}}\Rtil(t, \cb^*)$ 
	\end{tabular}
	\caption{\label{fig:ANN1} Method I: ANN approximations (blue points) for states and derivatives, compared to respective numerical solutions (red lines). Note the ANN approximations' slight departure from the numerical solutions, especially evident in the plots for $\frac{\pa}{\pa c_{I_0}}\Itil(t, \cb^*)$, $\frac{\pa}{\pa c_{I_0}}\Rtil(t, \cb^*)$, $\frac{\pa}{\pa c_{\gamma}}\Itil(t, \cb^*)$ and $\frac{\pa}{\pa c_{\beta}}\Rtil(t, \cb^*)$.}
\end{figure}

\begin{figure}
	\centering
	\begin{tabular}{ccc}
		\includegraphics[width = 5cm]{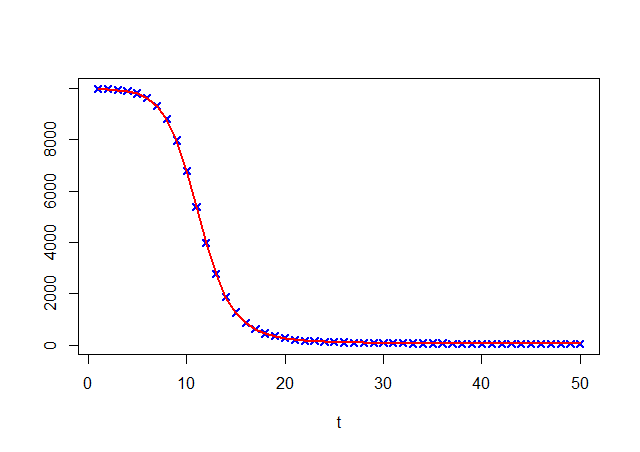} &
		\includegraphics[width = 5cm]{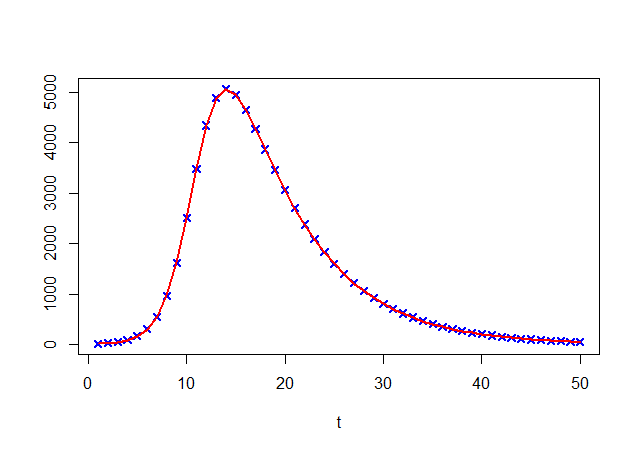} &
		\includegraphics[width = 5cm]{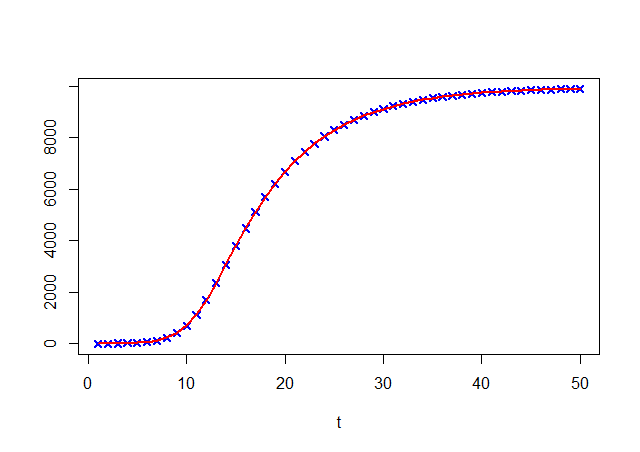} \\
		$\Stil(t, \cb^*)$ & $\Itil(t, \cb^*)$ & $\Rtil(t, \cb^*)$ \\[20pt]
		\includegraphics[width = 5cm]{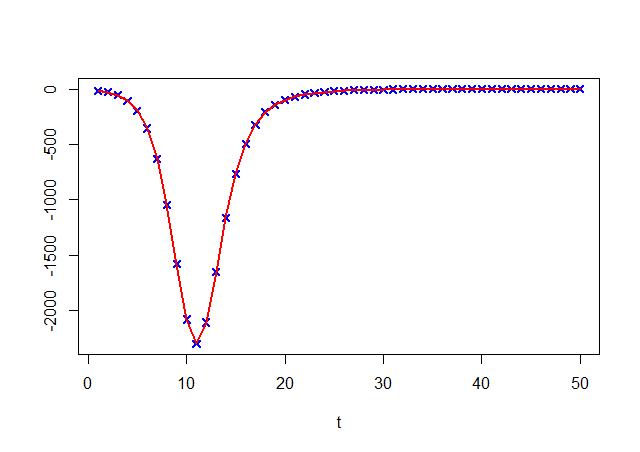} &
		\includegraphics[width = 5cm]{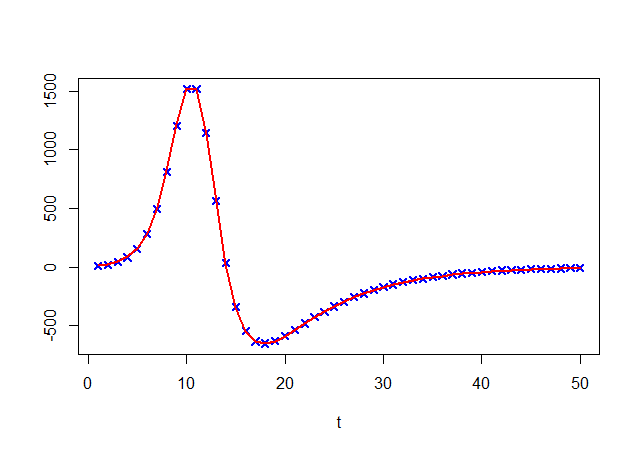} &
		\includegraphics[width = 5cm]{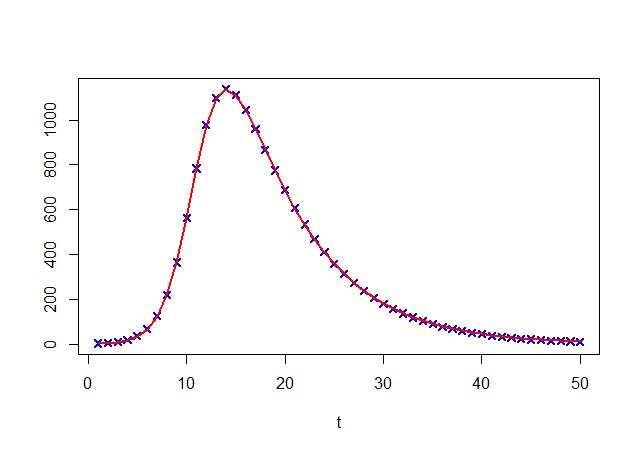} \\
		$\tilde{\frac{\pa S}{\pa c_{I_0}}}(t, \cb^*)$ & $\tilde{\frac{\pa I}{\pa c_{I_0}}}(t, \cb^*)$ & $\tilde{\frac{\pa R}{\pa c_{I_0}}}(t, \cb^*)$ \\[20pt]
		\includegraphics[width = 5cm]{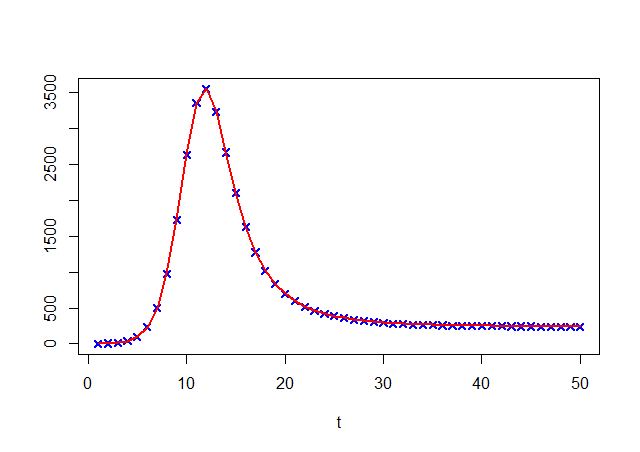} &
		\includegraphics[width = 5cm]{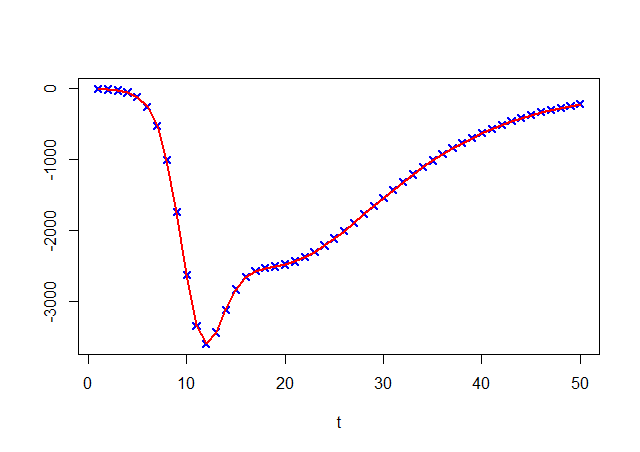} &
		\includegraphics[width = 5cm]{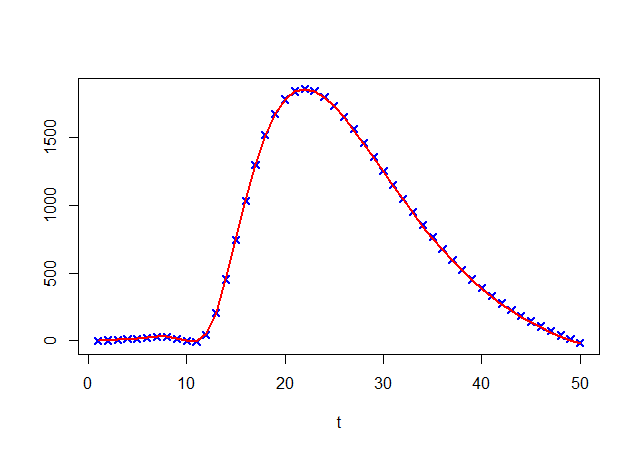} \\
		$\tilde{\frac{\pa S}{\pa c_{\gamma}}}(t, \cb^*)$ & $\tilde{\frac{\pa I}{\pa c_{\gamma}}}(t, \cb^*)$ & $\tilde{\frac{\pa R}{\pa c_{\gamma}}}(t, \cb^*)$ \\[20pt]
		\includegraphics[width = 5cm]{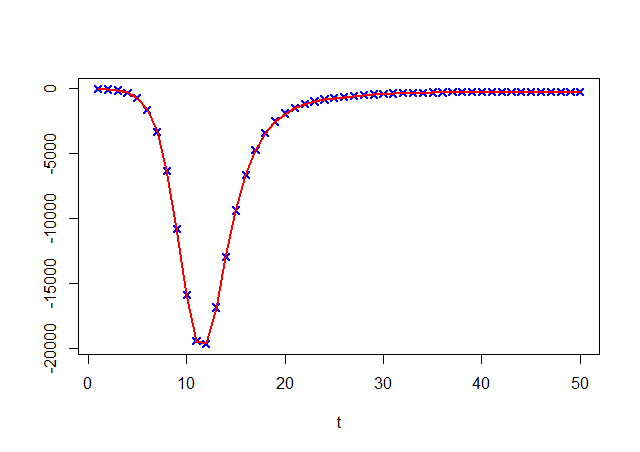} & 
		\includegraphics[width = 5cm]{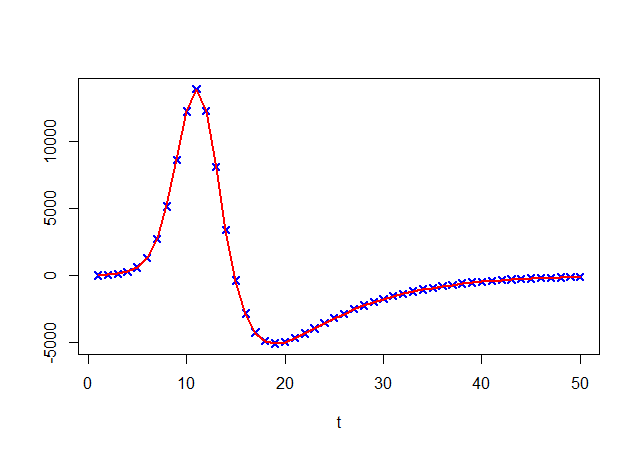} &
		\includegraphics[width = 5cm]{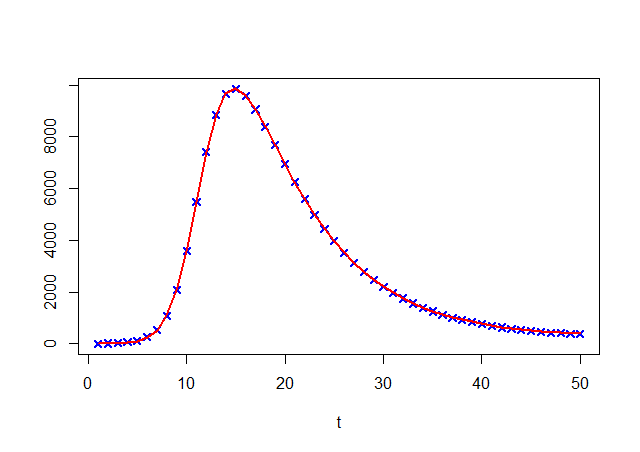} \\
		$\tilde{\frac{\pa S}{\pa c_{\beta}}}(t, \cb^*)$ & $\tilde{\frac{\pa I}{\pa c_{\beta}}}(t, \cb^*)$ & $\tilde{\frac{\pa R}{\pa c_{\beta}}}(t, \cb^*)$ \\
	\end{tabular}
	\caption{\label{fig:ANN2} Method II: ANN approximations (blue points) for states and derivatives, compared to respective numerical solutions (red lines). There are visible improvements in approximation abilities compared to Figure~\ref{fig:ANN1}; the departures from numerical solutions in Method II are very limited.}
\end{figure}

\renewcommand{\arraystretch}{0.75}
	
\begin{table}
	\begin{center}
		\begin{tabular}{|l|l|l|}
			\hline {States/Derivatives}
			& {Value of $p$ under Method I (\%)} & {Value of $p$ under Method II (\%)} \\
			\hline
			$S$ & 99.9992 & 99.9999 \\
			$I$ & 99.9984 & 99.9996 \\
			$R$ & 99.9990 & 99.9999 \\ \hline
			$\pa S/\pa c_{I_0}$ & 99.8749 & 99.9990 \\
			$\pa I/\pa c_{I_0}$ & 99.7152 & 99.9885 \\
			$\pa R/\pa c_{I_0}$ & 98.6116 & 99.9981 \\
			$\pa S/\pa c_{\gamma}$ & 99.9446 & 99.9977 \\
			$\pa I/\pa c_{\gamma}$ & 99.8515 & 99.9973 \\
			$\pa R/\pa c_{\gamma}$ & 99.7319 & 99.9919 \\
			$\pa S/\pa c_{\beta}$ & 99.9885 & 99.9964 \\
			$\pa I/\pa c_{\beta}$ & 99.9548 & 99.9928 \\
			$\pa R/\pa c_{\beta}$ & 99.9196 & 99.9976 \\ \hline
			$\pa^2 S/\pa c_{I_0}^2$ & 91.7210 & 99.9904 \\
			$\pa^2 I/\pa c_{I_0}^2$ & 95.1070 & 99.6717 \\
			$\pa^2 R/\pa c_{I_0}^2$ & 96.6019 & 99.1026 \\
			$\pa^2 S/\pa c_{I_0} \pa c_\gamma$ & 94.5013 & 99.9692 \\
			$\pa^2 I/\pa c_{I_0} \pa c_\gamma$ & 91.2081 & 99.8453 \\
			$\pa^2 R/\pa c_{I_0} \pa c_\gamma$ & 97.7580 & 99.9578 \\
			$\pa^2 S/\pa c_{I_0} \pa c_\beta$ & 99.4895 & 99.9906 \\
			$\pa^2 I/\pa c_{I_0} \pa c_\beta$ & 98.7981 & 99.9753 \\
			$\pa^2 R/\pa c_{I_0} \pa c_\beta$ & 95.7825 & 99.9674 \\
			$\pa^2 S/\pa c_{\gamma}^2$ & 75.3037 & 99.9737 \\
			$\pa^2 I/\pa c_{\gamma}^2$ & 92.4982 & 99.9271 \\
			$\pa^2 R/\pa c_{\gamma}^2$ & 91.4163 & 99.9206 \\
			$\pa^2 S/\pa c_{\gamma} \pa c_\beta$ & 89.8415 & 99.9793 \\
			$\pa^2 I/\pa c_{\gamma} \pa c_\beta$ & 90.1397 & 99.9415 \\
			$\pa^2 R/\pa c_{\gamma} \pa c_\beta$ & 97.7636 & 99.9541 \\
			$\pa^2 S/\pa c_{\beta}^2$ & 99.6090 & 99.9929 \\
			$\pa^2 I/\pa c_{\beta}^2$ & 98.2125 & 99.9627 \\
			$\pa^2 R/\pa c_{\beta}^2$ & 96.5849 & 99.9781 \\ \hline
		\end{tabular}
	\end{center}
	\caption{Numerical values for the percentage reduction from TSS for all states, first derivatives and second derivatives achieved by the respective ANNs at $\cb^*$, for Methods I and II.}
	\label{tab:sse}
\end{table}

Figure~\ref{fig:simposts} shows the marginal posterior distributions obtained for each parameter $c_{I_0}, c_\gamma, c_\beta$ from Method I, Method II and the MH algorithm, for each of the 200 simulated datasets. The posterior densities for all methods generally fall in the same area with little deviation; a more informative comparison is provided in Table~\ref{tab:mise} detailing the approximate Mean Integrated Squared Errors (MISE) \cite{wand1994kernel} between the two posterior density estimates from each of our methods and that from the MH algorithm calculated using 
\begin{align*}
	MISE_{\phi}(h_1, h_2; \yb) & := \frac{1}{200} \sum_{i=1}^{200} \int_{-\infty}^{\infty} (h_1(\phi; \yb_i) - h_2(\phi; \yb_i))^2 \ d\phi \\
	& \approx \frac{1}{200} \sum_{i=1}^{200} \sum_{j = 1}^{H} {\left(h_1(\phi^{(j)}) - h_2(\phi^{(j)})\right)^2 \left(\phi^{(j+1)} - \phi^{(j)}\right)},
	\numberthis \label{eq:mise}
\end{align*}
for $\phi \in \{c_{I_0}, c_\gamma, c_\beta\}$, where $h_1, h_2$ are densities to be compared, and $\{\phi^{(j)}\}_{j = 1, ..., H}$ are equally spaced points with $[\phi^{(1)}, \phi^{(H+1)}]$ chosen to cover a large enough range of values of $\phi$ and $H$ large to ensure good approximation to the true MISE. Hence Table~\ref{tab:mise} shows that Method II gives posteriors closer to the MH algorithm overall. The pointwise average of the 200 posterior densities are also plotted in Figure~\ref{fig:simpost}.

\begin{figure}
	\centering
	\begin{tabular}{ccc}
		\includegraphics[width = 5cm]{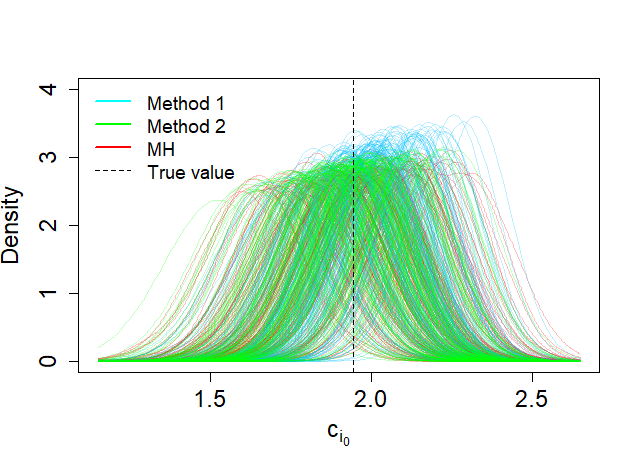} &
		\includegraphics[width = 5cm]{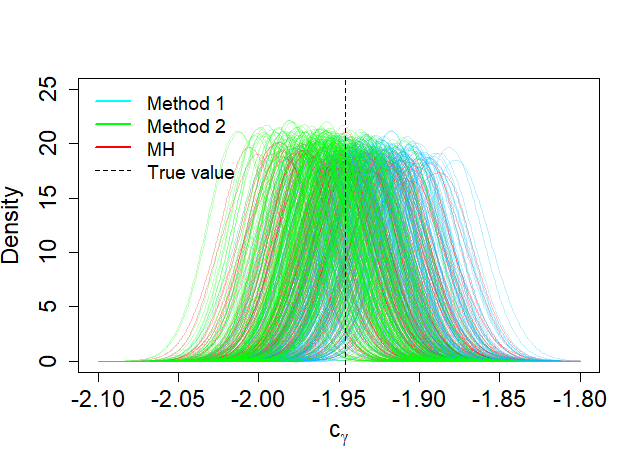} &
		\includegraphics[width = 5cm]{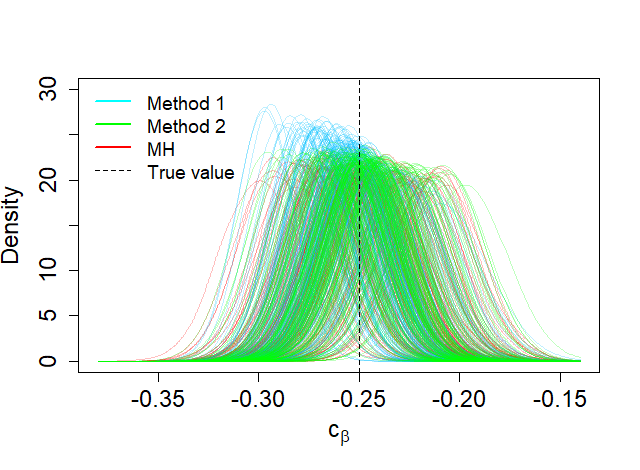} \\
		(a) & (b) & (c)
	\end{tabular}
	\caption{\label{fig:simposts} Variability of the marginal posterior distributions for the 200 simulated datasets (Blue: Method I, Green: Method II, Red: MH algorithm, Black line: True value) corresponding to the parameters (a) $c_{I_0}$, (b) $c_{I_0}$ and (c) $c_\beta$.}
\end{figure}

\begin{table}
	\begin{center}
		\begin{tabular}{|l|l|l|}
			\hline {} & \multicolumn{2}{c|}{MISE} \\
			\hline {Parameter}
			& {MH vs. Method I} & {MH vs. Method II} \\
			\hline
			$c_{I_0}$ & 0.106 & 0.0393 \\
			$c_\gamma$ & 2.32 & 2.59 \\
			$c_\beta$ & 0.706 & 0.307 \\
			\hline
		\end{tabular}
	\end{center}
	\caption{\label{tab:mise} Numerical values of MISE between the marginal posterior distributions corresponding to the parameters $c_{I_0}, c_\gamma$ and $c_\beta$ in the simulated datasets example.}
\end{table}

\begin{figure}
	\centering
	\begin{tabular}{ccc}
		\includegraphics[width = 5cm]{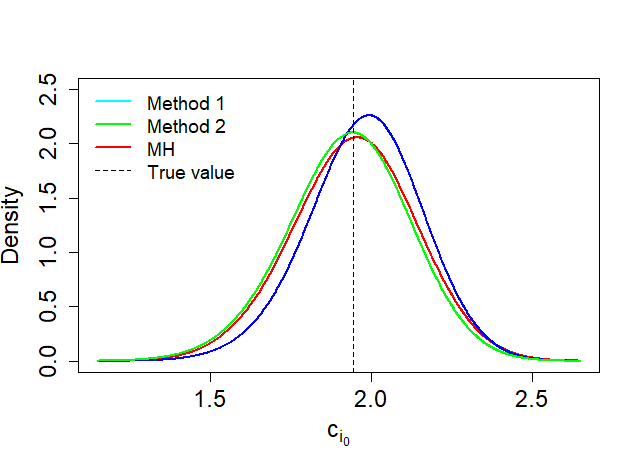} &
		\includegraphics[width = 5cm]{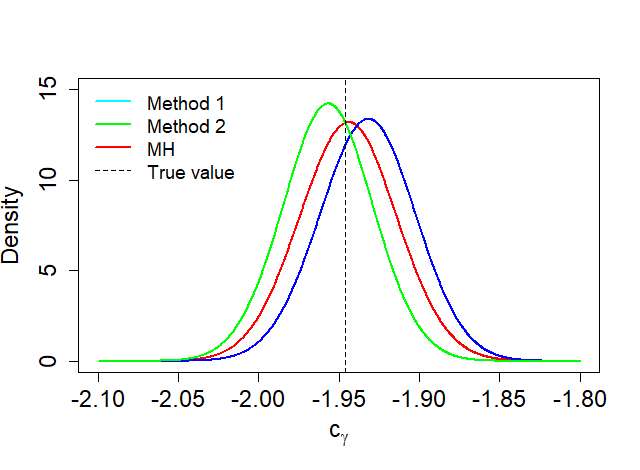} &
		\includegraphics[width = 5cm]{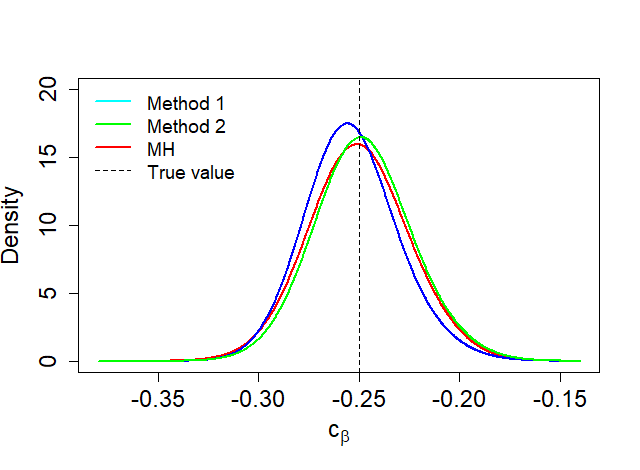} \\
		(a) & (b) & (c)
	\end{tabular}
	\caption{\label{fig:simpost} Point-wise average of marginal posterior distributions based on the 200 simulated datasets (Blue: Method I, Green: Method II, Red: MH algorithm, Black line: True value) corresponding to the parameters (a) $c_{I_0}$, (b) $c_{I_0}$ and (c) $c_\beta$.}
\end{figure}

\section{Influenza Example: Real Data}
\label{section:influenza}

To demonstrate the real-life applicability of our proposed methods, we infer the parameters $c_{I_0}, c_\gamma, c_\beta$ from an influenza outbreak at a boys boarding school reported by the Lancet \cite{influenza1978} in 1978, shown in Table~\ref{tab:influenzadata}. This epidemic dataset provides the number of individuals confined to bed over time and so corresponds to the infectious state ($I$) in the SIR model. Raissi \cite{raissi2019parameter} also applied their physics-informed deep learning methodology onto this dataset to infer the values of $\beta$ and $\gamma$. Similar to Section~\ref{section:methods}, we will outline the training and design of the relevant ANNs, as well as the priors, likelihoods and posteriors pertaining to the Laplace Approximations that follow. Lastly, the results from Methods I and II will also be compared to that obtained from a benchmark MH algorithm.

\renewcommand{\arraystretch}{0.75}

\begin{table}
	\begin{center}
		\begin{tabular}{|l|l|l|l|}
			\hline 
			Day & Infected Cases & Day & Infected Cases\\
			\hline
			1 & 3 & 8 & 235 \\
			2 & 8 & 9 & 190 \\
			3 & 28 & 10 & 126 \\
			4 & 75 & 11 & 70 \\
			5 & 221 & 12 & 28 \\
			6 & 291 & 13 & 12 \\
			7 & 255 & 14 & 5 \\
			\hline
		\end{tabular}
	\end{center}
	\caption{\label{tab:influenzadata} Boys School influenza dataset \cite{influenza1978}.}
\end{table}

\renewcommand{\arraystretch}{1}

\subsection{ANN: Training Examples and Architecture}
\label{subs:infann}

A challenge present in inferring the ODE parameters from real-life datasets using our methods is the choice of hyperparameters for the ANN's collocation grid; MCMC algorithms face a somewhat similar challenge in the choice of initial points. We could specify a large range of values to hopefully include the true value of the parameters, but if the grid density is to be kept reasonably high (to ensure good approximation between collocation points), the number of training examples would become extremely large. In light of this, we use a simpler, preliminary Maximum Likelihood Estimation algorithm to narrow down the range of values to be included in our collocation grid. \\

The approximate Maximum Likelihood Estimation algorithm uses a Monte Carlo approach as follows. First, 20,000 sets of parameters are generated from a large range of values, each of which will be used to numerically solve the SIR ODE system based on the observed Influenza data. The likelihood under a Poisson model is calculated for each set of parameters, and the set with the largest likelihood is chosen as the (approximate) Maximum Likelihood Estimate (MLE). This procedure results in an approximation to the true MLE which increases in accuracy as the number of sets of parameters generated becomes large. However, for the purpose of grid construction only, this approximation approach suffices for training the ANNs within bounds of reasonable parameter values, and takes less than a minute. The values we obtained from this algorithm were $(-0.574, -0.704, 0.593) = (\ln 0.563, \ln 0.495, \ln 1.810)$. This inspires the following choice of hyperparameters for the collocation grid:
\begin{align*}
	& t_1^C = 1, \ && t_{N_t}^C = 14, \ && N_t = 14; \\
	& c_{I_0, 1}^C = \ln 0.1, \ && c_{I_0, N_{c_{I_0}}}^C = \ln 3, \ && N_{c_{I_0}} = 31; \\
	& c_{\gamma, 1}^C = \ln 1/3, \ && c_{\gamma, N_{c_\gamma}}^C = \ln 1/1, \ && N_{c_\gamma} = 11; \\
	& c_{\beta, 1}^C = 0.4, \ && c_{\beta, N_{c_\beta}}^C = 0.8, \ && N_{c_\beta} = 6.
	\numberthis
	\label{eq:influenzagrid}
\end{align*} 

Before obtaining the training targets, we carried out an additional step to exclude parameter combinations that admitted a basic reproduction number $R_0 := e^{c_\beta - c_\gamma}$ of less than 1, since $R_0 < 1$ leads to a decline in the number of infected individuals over the entire time frame, which is clearly not the case in this dataset; in fact, this step removes unnecessary training examples and reduces computation time. \\

Subsequently, training targets were obtained similar to Section~\ref{subs:dnnarch2} - numerically solving the ODE for all parameter combinations and then applying $z$-score standardizations. The depths and widths chosen for each ANN were $K = 2$ and $L_1 = L_2 = 10$ with $tanh$ activations. $J_1$ and $J_2$ were used as loss functions for Methods I and II respectively, however we found some different optimal values for the regularization coefficients: $\lambdab_{11} = (0.5, 0.5, 0.5), \lambdab_{12} = 0.5, \lambda_{21} = (0.1, 0.1, 0.1, 0.1, 0.1, 0.1,0.1, 0.1, 0.1,0.1, 0.1, 0.1)$ and $\lambdab_{22} = (0.1, 0.1, 0.1, 0.1)$.

\subsection{Laplace Approximation}

The same vague priors as in Section~\ref{subs:2laplace} were chosen for $\cb$; one could use a more informative prior based on the approximate MLE for a more reliable inference, but in this case study, a vague prior choice was sufficient. \\

Modelling the observed infected cases $\yb := (y(t_1), ..., y(t_{14}))$ after a Poisson distribution with mean $\Itil(t, \cb)$, the likelihood is expressed as:
\begin{align*}
	\mathcal{L}(\cb; \yb) = \prod_{n=1}^{14} p(y(t_n)|\cb) \approx \prod_{n=1}^{14} \text{Poi}\left(y(t_n)\middle|\Itil(t_n, \cb)\right) = 
	\prod_{n=1}^{14} \frac
	{\Itil(t_n, \cb)^{y(t_n)} e^{-\Itil(t_n, \cb)}}
	{y(t_n)!}
\end{align*}
which gives the log-likelihood
\begin{align*}
	\ell(\cb, \yb) \approx \sum_{n=1}^{14} y(t_n) \ln \Itil(t_n, \cb) - \sum_{n=1}^{14} \Itil(t_n, \cb) + \text{const. indep. of } \cb.
\end{align*}

The corresponding expression for the function $g$ in this real-life example is then:
\begin{align*}
	g(\cb) = \ln p(\cb) + \ln p(\yb|\cb) = -\frac{1}{2\times 10^4}(c_{I_0}^2 + c_\gamma^2 + c_\beta^2) + \sum_{n=1}^{N} y(t_n) \ln \Itil(t_n, \cb) - \sum_{n=1}^{N} \Itil(t_n, \cb).
\end{align*}

The BFGS algorithm with an initial point at the approximate MLE $(-0.574, -0.704, 0.593)$ was then used to find $\hat{\cb}$, and subsequently, $\nabla^{-2} g(\hat{\cb})$, which determine the approximate posterior $h(\cb|\yb) \sim N(\hat{\cb}, \nabla^{-2} g(\hat{\cb}))$.

\subsection{Performance Benchmark}

A Metropolis-Hastings algorithm similar to Section~\ref{subs:dnnmh} was carried out, with the main difference being the likelihood $p(\yb|\cb) = \prod_{n=1}^{14} \text{Poi}(y(t_n)|I(t_n, \cb))$ taken as the Poisson distribution. The hyperparameters $n_{iter}$ and $\alpha$ were chosen to be $200,000$ and $1,000$, with $s_{c_{I_0}}^2 = s_{c_\gamma}^2 = s_{c_\beta}^2 = 0.05$. A kernel density estimate with appropriate bandwidth was applied to the $2,000$ posterior samples from the MH algorithm to be compared with $h(\cb|\yb)$ from Methods I and II.

\subsection{Results}
\label{subs:infresults}

Table~\ref{tab:infsse} shows the percentage reduction $p$ from the total sum of squares (TSS) by the ANNs in Methods I and II at the approximate MLE $(-0.574, -0.704, 0.593)$; Method II performs comparably with Method I in approximating the states and first derivatives, but fares much better in approximating the second derivatives.\\

\renewcommand{\arraystretch}{0.75}

\begin{table}
	\begin{center}
		\begin{tabular}{|l|l|l|}
			\hline {States/Derivatives}
			& {Value of $p$ under Method I (\%)} & {Value of $p$ under Method II (\%)} \\
			\hline
			$S$ & 99.9974 & 99.9975 \\
			$I$ & 99.9943 & 99.9947 \\
			$R$ & 99.9989 & 99.9960 \\ \hline
			$\pa S/\pa c_{I_0}$ & 99.9798 & 99.9919 \\
			$\pa I/\pa c_{I_0}$ & 99.9525 & 99.9946 \\
			$\pa R/\pa c_{I_0}$ & 99.9529 & 99.9851 \\
			$\pa S/\pa c_{\gamma}$ & 99.9768 & 99.9538 \\
			$\pa I/\pa c_{\gamma}$ & 99.8068 & 99.9677 \\
			$\pa R/\pa c_{\gamma}$ & 98.4257 & 99.4891 \\
			$\pa S/\pa c_{\beta}$ & 99.9954 & 99.9958 \\
			$\pa I/\pa c_{\beta}$ & 99.9175 & 99.9958 \\
			$\pa R/\pa c_{\beta}$ & 99.9416 & 99.9940 \\ \hline
			$\pa^2 S/\pa c_{I_0}^2$ & 99.9017 & 99.9738 \\
			$\pa^2 I/\pa c_{I_0}^2$ & 99.4937 & 99.9642 \\
			$\pa^2 R/\pa c_{I_0}^2$ & 99.2695 & 99.9622 \\
			$\pa^2 S/\pa c_{I_0} \pa c_\gamma$ & 99.9206 & 99.9647 \\
			$\pa^2 I/\pa c_{I_0} \pa c_\gamma$ & 99.3422 & 99.8975 \\
			$\pa^2 R/\pa c_{I_0} \pa c_\gamma$ & 96.7491 & 99.8389 \\
			$\pa^2 S/\pa c_{I_0} \pa c_\beta$ & 99.9174 & 99.9762 \\
			$\pa^2 I/\pa c_{I_0} \pa c_\beta$ & 99.5373 & 99.9678 \\
			$\pa^2 R/\pa c_{I_0} \pa c_\beta$ & 99.7958 & 99.9716 \\
			$\pa^2 S/\pa c_{\gamma}^2$ & 99.6439 & 99.9178 \\
			$\pa^2 I/\pa c_{\gamma}^2$ & 95.0311 & 99.8142 \\
			$\pa^2 R/\pa c_{\gamma}^2$ & 98.6503 & 99.9469 \\
			$\pa^2 S/\pa c_{\gamma} \pa c_\beta$ & 99.9361 & 99.9233 \\
			$\pa^2 I/\pa c_{\gamma} \pa c_\beta$ & 98.1427 & 99.6939 \\
			$\pa^2 R/\pa c_{\gamma} \pa c_\beta$ & 98.1007 & 99.7795 \\
			$\pa^2 S/\pa c_{\beta}^2$ & 99.8928 & 99.9715 \\
			$\pa^2 I/\pa c_{\beta}^2$ & 98.7511 & 99.9769 \\
			$\pa^2 R/\pa c_{\beta}^2$ & 99.0303 & 99.9712 \\ \hline
		\end{tabular}
	\end{center}
	\caption{Numerical values for the percentage reduction from TSS for all states, first derivatives and second derivatives achieved by the respective ANNs at the approximate MLE $(-0.574, -0.704, 0.593)$, for Methods I and II.}
	\label{tab:infsse}
\end{table}

Figure~\ref{fig:infpost} shows the resulting approximate posteriors $h(\cb|\yb)$ for both Methods I and II, plotted together with the kernel density estimate from the benchmark MH algorithm. Method I yielded the approximate marginal posteriors for $c_{I_0}, c_\gamma$ and $c_\beta$ which are, respectively, 
\begin{align*}
	& c_{I_0}|\yb \sim N(-0.837, 0.0495), \\
	& c_{\gamma}|\yb \sim N(-0.730, 5.3249\times 10^{-4}), \text{ and}\\
	& c_{\beta}|\yb \sim N(0.619, 6.7071\times 10^{-4}),
	\numberthis
	\label{eq:infpost1}
\end{align*} whereas Method II yielded 
\begin{align*}
	& c_{I_0}|\yb \sim N(-0.853, 0.0619), \\
	& c_{\gamma}|\yb \sim N(-0.729, 5.5937\times 10^{-4}), \text{ and}\\
	& c_{\beta}|\yb \sim N(0.621, 8.0506\times 10^{-4}).
	\numberthis
	\label{eq:infpost2}
\end{align*} 
The MH algorithm resulted in the following marginal posterior samples: the MAP estimates and posterior sample variances were $(-0.944, 0.0633)$ for $c_{I_0}$, $(-0.730, 5.5121\times 10^{-4})$ for $c_\gamma$ and $(0.630, 7.9948\times 10^{-4})$ for $c_\beta$, respectively; note that the density estimates from the MH algorithm are not necessarily Gaussian. The approximate Integrated Squared Error (ISE) 
\begin{align*}
	ISE_\phi(h_1, h_2; \yb) & := \int_{-\infty}^{\infty} (h_1(\phi; \yb_i) - h_2(\phi; \yb_i))^2 \ d\phi \\
	& \approx \sum_{j = 1}^{H} {\left(h_1(\phi^{(j)}) - h_2(\phi^{(j)})\right)^2 \left(\phi^{(j+1)} - \phi^{(j)}\right)},
	\numberthis \label{eq:ise}
\end{align*}
between the posterior density from each of our proposed methods ($h_1$) and the kernel density estimate from the MH algorithm ($h_2$) are displayed in Table~\ref{tab:infmise}. \\

\begin{figure}
	\centering
	\begin{tabular}{ccc}
		\includegraphics[width = 5cm]{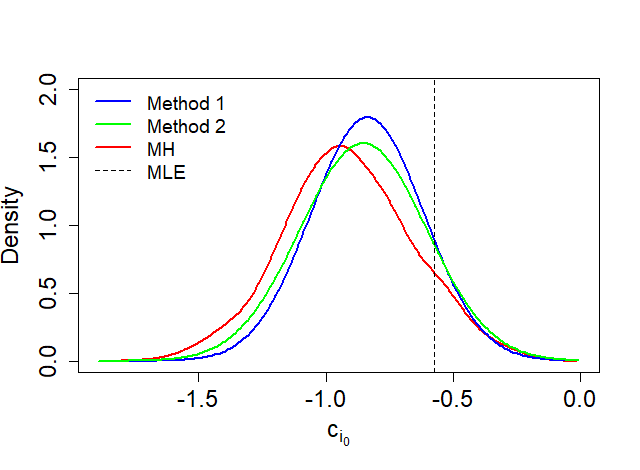} &
		\includegraphics[width = 5cm]{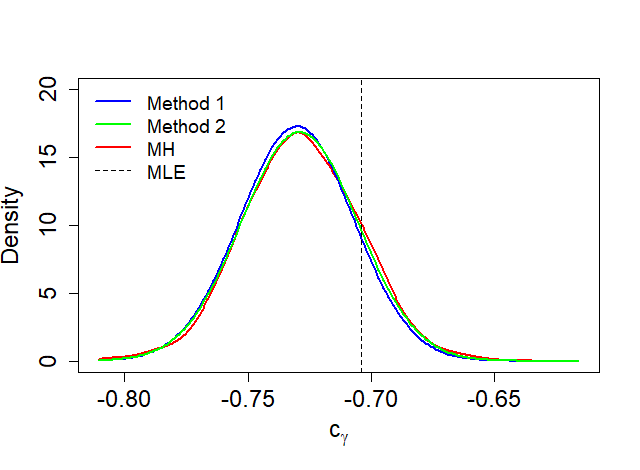} &
		\includegraphics[width = 5cm]{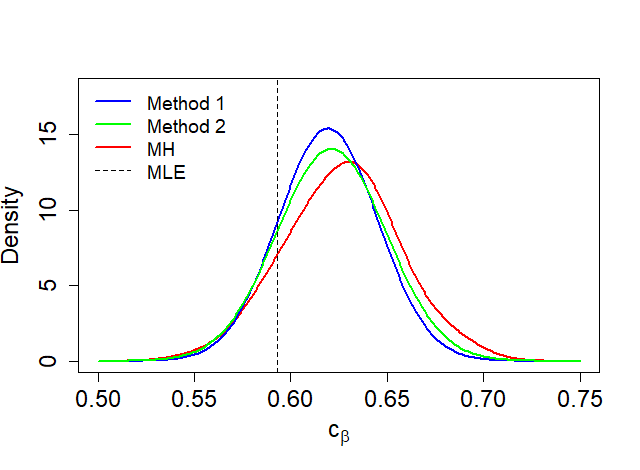} \\
		(a) & (b) & (c) \\
	\end{tabular}
	\caption{\label{fig:infpost} Marginal posterior distributions for (a) $c_{I_0}$ (b) $c_\gamma$ and (c) $c_\beta$ for the influenza dataset (Blue: Method I, Green: Method II, Red: MH algorithm, Black line: approximate MLE).}
\end{figure}

\begin{table}
	\begin{center}
		\begin{tabular}{|l|l|l|}
			\hline {} & \multicolumn{2}{c|}{ISE} \\
			\hline {Parameter}
			& {MH vs. Method I} & {MH vs. Method II} \\
			\hline
			$c_{I_0}$ & 0.0962 & 0.0438 \\
			$c_\gamma$ & 0.0534 & 0.0134 \\
			$c_\beta$ & 0.548 & 0.234 \\
			\hline
		\end{tabular}
	\end{center}
	\caption{\label{tab:infmise} ISE between the posterior density estimate from the MH algorithm and the approximate posteriors from each of Methods I and II for $c_{I_0}, c_\gamma$ and $c_\beta$ in the influenza example.}
\end{table}

In summary, transforming the parameters back to the original scales (instead of the logarithmic scale), the MAP estimates for $(I(0), \gamma, \beta)$, $(0.433, 0.482, 1.858)$, $(0.426, 0.482, \\ 1.861)$ and $(0.389, 0.482, 1.877)$ in Method I, Method II and the benchmark MH algorithm respectively. Since $I(0)$ takes discrete values in reality, we conclude that the most likely value of $I(0)$ is 1 for both methods, by virtue of the posteriors found. 

\subsection{Additional Investigations}
\label{subs:addinv}

In this subsection, we investigate the posterior inference under two additional settings: (i) when the initial point for the BFGS algorithm is further from the optimum, and (ii) when the prior for $c_{I_0}$ is specified to be very informative (low variance) and centered around $0$ (i.e. $I(0) = 1$) in light of the conclusion drawn in the previous section. \\

Using the same ANN trained in Section~\ref{subs:infann}, we repeat the posterior inference with a different initial point $(\ln 2.9, \ln 1/2.9, 0.41)$ (near the boundaries of the collocation grid) for the BFGS algorithm. Method I results in the following approximate marginal posteriors for $c_{I_0}, c_\gamma$ and $c_\beta$ given by:
\begin{align*}
	& c_{I_0}|\yb \sim N(-0.838, 0.0495), \\
	& c_{\gamma}|\yb \sim N(-0.730, 5.3248 \times 10^{-4}), \text{ and} \\
	& c_{\beta}|\yb \sim N(0.619, 6.7090 \times 10^{-4}),
\end{align*}
whereas Method II gives:
\begin{align*}
	& c_{I_0}|\yb \sim N(-0.853, 0.0619), \\
	& c_{\gamma}|\yb \sim N(-0.729, 5.5937 \times 10^{-4}), \text{ and} \\
	& c_{\beta}|\yb \sim N(0.621, 8.0506 \times 10^{-4}), \text{ respectively,}
\end{align*}
all of which are very close to the previous estimates in Equations (\ref{eq:infpost1}) and (\ref{eq:infpost2}). Notably, the MAP and posterior variance estimates are identical up to at least 4 decimal places for Method II. \\

To fix $I(0) = 1$, we construct another ANN that only takes $c_\gamma$ and $c_\beta$ as inputs. Naturally, the outputs for Method II and the loss functions will only involve derivatives with respect to $c_\gamma$ and $c_\beta$, but all other hyperparameters are kept the same. The resulting approximate marginal posteriors for $c_\gamma$ and $c_\beta$ from Method I are:
\begin{align*}
	& c_{\gamma}|\yb \sim N(-0.737, 5.3980 \times 10^{-4}), \text{ and}\\
	& c_{\beta}|\yb \sim N(0.525, 8.3290 \times 10^{-5}),
\end{align*}
and those for Method II are:
\begin{align*}
	& c_{\gamma}|\yb \sim N(-0.735, 5.8574 \times 10^{-4}), \text{ and}\\
	& c_{\beta}|\yb \sim N(0.527, 8.6412 \times 10^{-5}), \text{ respectively,}
\end{align*}
and a comparison with the corresponding kernel density from MH algorithm for the inference of $c_\gamma$ and $c_\beta$ is plotted in Figure~\ref{fig:infpostfixedi0}. Numerically, the MAP estimates and variances obtained from the corresponding MH algorithm are $(-0.732, 5.5445\times 10^{-4})$ for $c_\gamma$ and $(0.5238, 8.3951\times 10^{-5})$ for $c_\beta$, respectively.  Alternatively, using a very small prior variance for $c_{I_0}$ in the original algorithm for $\cb$ yields similar results.

\begin{figure}
	\centering
	\begin{tabular}{cc}
		\includegraphics[width = 5cm]{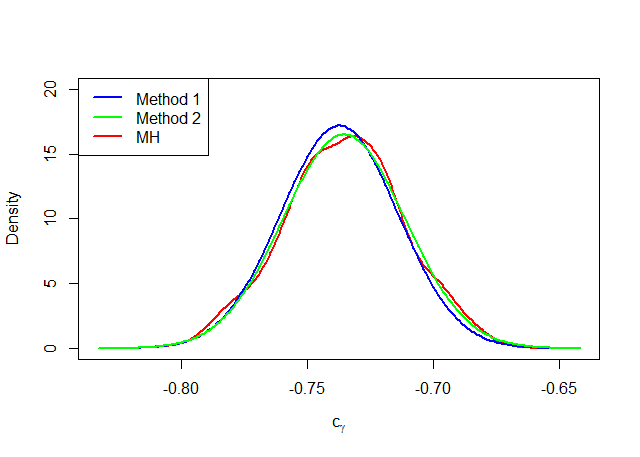} &
		\includegraphics[width = 5cm]{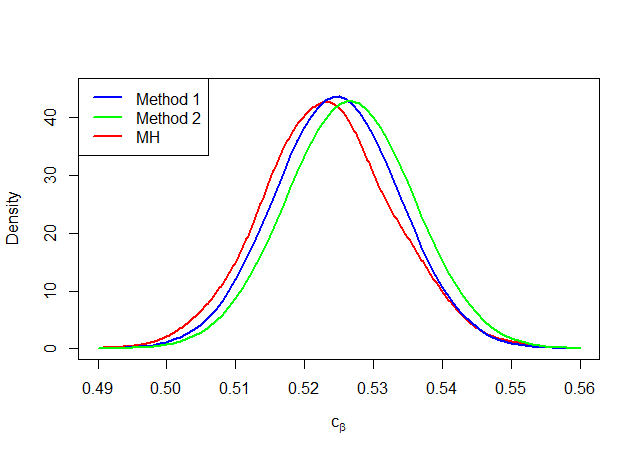} \\
		(a) & (b)
	\end{tabular}
	\caption{\label{fig:infpostfixedi0} Marginal posterior distributions for (a) $c_\gamma$ and (b) $c_\beta$ (Blue: Method I, Green: Method II, Red: MH algorithm) for the influenza example, with fixed $I(0) = 1$.}
\end{figure}

\section{Discussion}
\label{section:discussion}

\subsection{Method I vs. Method II}
\label{subs:4comp12}

The difference of Method II from Method I lies in the training of the derivatives $\nabla_{\phib} \xb$ as well as the regularization terms pertaining to these derivatives. In the simulated example, the ANN from Method II did better in approximating all states and derivatives compared to Method I (see Table~\ref{tab:sse}). In the real-life example, the ANN in Method II performed better in approximating second derivatives, but was comparable to Method I in approximating $S, I, R$ and their first derivatives (see Table~\ref{tab:infsse}); this is still intuitively plausible since Method II does not perform worse than Method I overall - the extra outputs and regularizations still contributed to an improvement in approximations, even if only marginal. The discrepancies between the former and the latter may have several explanations, among which are: 
\begin{itemize}
	\item Random initializations of the ANN weights and biases $w_{ij}^{(k)}, b_j^{(k)}$: the loss functions proposed could have multiple minima, especially for the SIR ODE system which can be rather complex. Slight differences in initialization could steer the ANN towards different minima and result in different weights and biases;
	\item Accuracy tradeoffs due to training more states in the same number of epochs: Both ANNs in Methods I and II were trained for 2,500 epochs, but the ANN in Method II has a concatenated output of a larger dimension - this means the optimization process is more complex in Method II, and a larger number of epochs may be needed to achieve consistently better predictions for $S, I, R$ and their first derivatives;
	\item Choice of regularization coefficients: $\lambdab_{11}, \lambdab_{12}, \lambdab_{21}, \lambdab_{22}$ were determined via trial and error in both Methods I and II. Due to the amount of time consumed to train an ANN, only a limited number of combinations could be examined. Even though we have found that the performance of the ANN does not vary largely within a small neighbourhood of the optimal values, it could explain the small discrepancy between the simulated dataset ANNs and the influenza example ANNs. Perhaps a more thorough optimization algorithm for the regularization coefficients would lead to Method II performing consistently better than Method I, but that is beyond the scope of this paper.
\end{itemize} 

In estimating the posterior distributions, we see in Figures~\ref{fig:simpost} and \ref{fig:infpost}, as well as in Tables~\ref{tab:mise} and \ref{tab:infmise}, that Method II produces posterior distributions closer to that obtained with the benchmark MH algorithm, compared to Method I in both simulated datasets and influenza examples. This demonstrates the two-fold advantage that Method II provides over Method I: more accurate approximations to the first and second derivatives, leading to a more accurate MAPs found via the gradient BFGS algorithm and better uncertainty quantifications respectively. This does, however, come with heavier computational costs - the time taken to train each ANN is shown in Table~\ref{tab:traintime}. The BFGS algorithm runtime for the Laplace Approximation is also shown in Table~\ref{tab:bfgstime} - this is faster under Method II since the first derivatives required for the algorithm are directly obtained from the ANN output, whereas differentiation is required to obtain these first derivatives under Method I.\\

\begin{table}
	\begin{center}
		\begin{tabular}{|l|l|l|}
			\hline {} & \multicolumn{2}{c|}{Time taken to train ANN } \\
			\hline {Example}
			& {Method I} & {Method II} \\
			\hline
			Simulated & 25.7 minutes & 1 hour, 28.0 minutes \\
			Influenza & 23.6 minutes & 38.3 minutes \\
			\hline
		\end{tabular}
	\end{center}
	\caption{\label{tab:traintime} Time taken to train ANNs for Methods I and II  on a HP workstation with 16 GB RAM and 12 Intel Core 7.0 processors with processing speed of 2.60 GHz, in the simulated dataset and influenza examples.}
\end{table}

\begin{table}
	\begin{center}
		\begin{tabular}{|l|l|l|l|}
			\hline {} & \multicolumn{3}{c|}{Time taken for posterior inference algorithm} \\
			\hline {Example}
			& {Method I} & {Method II} & {MH Algorithm} \\
			\hline
			200 Simulated Datsets & 1 hour, 32 minutes & 5.9 minutes & 39 hours, 41.8 minutes\\
			Influenza & 1.72 seconds & 1.85 seconds & 9.7 minutes \\
			\hline
		\end{tabular}
	\end{center}
	\caption{\label{tab:bfgstime} Time taken for posterior inference algorithms  under Methods I and II, on a HP workstation with 16 GB RAM and 12 Intel Core 7.0 processors with processing speed of 2.60 GHz, in the simulated dataset and influenza examples.}
\end{table}

Lastly, we also see from Section~\ref{subs:addinv} that the results are not sensitive to the initial point of the BFGS optimization algorithm - this is due to good approximations to the first and second dertivatives, which render the optimization algorithm and uncertainty quantification consistent, demonstrating the importance of sufficient training as well as regularization of the ANNs. Additional evidence of this is the near-identical inference of the posterior under Method II despite a relatively large change in the initial point. We also found that reducing the extent of training of the ANN increases the sensitivity of the MAP to the initial point, as expected.

\subsection{Proposed Methods vs. Random Walk MH}
\label{subs:dnnmh}

The benchmark MH algorithm utilizes the numerical solution to the ODE system to calculate likelihoods, and with a large number of iterations as well as good mixing, it produces a very representative sample of the posterior. The reliability of this algorithm lends itself naturally to be the benchmark for our proposed methods. As the number of iterations tends to infinity, the posterior distribution inferred using the MH algorithm represents the ceiling of the performance our methods, since our ANNs are also trained with numerical solutions as targets. \\

The main drawback of the MH algorithm in the inference of parameters of an analytically intractable ODE system is the computational cost - the ODE systems needs to be solved for every proposed move, causing a long run-time; see Table~\ref{tab:bfgstime}. For $n_{iter} = 200,000$ and $\alpha = 1,000$, the MH algorithm takes more than 39 hours to generate appropriate posterior samples for the 200 simulated datasets, significantly longer than the time taken for the Laplace approximation algorithm to produce the corresponding approximate posteriors in both Methods I and II; Method II especially produces results comparable to the MH algorithm. This highlights ANNs as a flexible function approximation tool which allow for quick evaluation of the approximate solution of the ODE system, and hence, a fast and convenient posterior inference, at the sacrifice of little accuracy. \\

Furthermore, the MH algorithm becomes inefficient when sampling in high dimensions especially when there are high correlations between parameters. In other general MCMC methods, this problem may be overcome by modifying the jump proposals or the acceptance criteria, potentially changing the space from which the algorithm is able to sample from. Inefficiencies in high dimensions may occur with ANNs as well, but can be circumvented using deeper and wider networks, or better optimizers that adapt to different difficulties in optimization; the ANN approach makes the optimization task explicit and independent so that specific modifications can be made only to the optimizer, without changing the structure or the function space that it may access.

\subsection{General Advantages and Limitations}

The inferential capabilities of the ANN approach coupled with the Laplace approximation are comparable to a typical MCMC approach, such as the Random Walk MH algorithm. Once trained, the ANN approximates the trajectory/solution surface at all points within the designed range, then carries out a relatively lightweight optimization task based on the Laplace Approximation to obtain the approximate posterior. This may prove convenient if the true values of the ODE parameters change to a nearby value within the designed range; the prediction and gradient surfaces remain accurate and can be re-used in the Laplace Approximation optimization. This is in contrast to the MH algorithm, where the entire algorithm needs to be run again to infer the corresponding posterior distribution, considerably increasing the computational cost, especially if the true values change frequently or sequentially. This advantage of ANNs over MCMC algorithms become more evident in sequential settings when the latter needs to be rerun every time a new observation arrives. \\

Another advantage of the ANN approach, along with other function approximation methods such as basis function expansion, stems from the flexibility of its loss or objective function, allowing regularization to be applied to improve posterior inference of the ODE parameters. The extent of regularization can be further controlled by varying the values of the regularization coefficients $\lambdab$, allowing a bias-variance tradeoff. Traditional MCMC methods lack this flexibility, and their hyperparameters often concern the mixing efficiency and sufficient representation of the posterior, rather than controlling the bias-variance tradeoff. \\

The main limitation to our approach is the lack of flexibility of the approximate posterior, since the Laplace Approximation restricts it to be a Gaussian density. Affine transformations performed on the parameters can overcome this restriction to some extent (much like the log-transformation on our original parameters $I(0), \gamma$ and $\beta$), but the posterior distribution would still be restrictied to only certain transformations of the Gaussian distribution. This is slightly evident in Figure~\ref{fig:infpost} where the posterior distribution for $c_\beta$ (and to a lesser extent, for $c_{I_0}$) resulting from the MH algorithm is slightly skewed, a characteristic that was not captured by Method I or II. Traditional MCMC approaches are not bound by this limitation and are flexible in sampling from any well-behaved posteriors. A consolation to our method is that it may be extended to use the approximate posterior in conjunction with importance sampling to obtain a better representation of the posterior, by employing a reweighing of the samples generated from the Laplace method. In that sense, the Laplace-approximated posterior can be seen as an effective way to approximate the true posterior since the resampling weights quickly deteriorate if the approximate is crude .\\

Methods I and II are based on design - particularly the grid of values we used to train the ANN, and therefore inherits the limitation we have imposed in the process of fixing this design. In our applications, we used a relatively small grid of values specified in (\ref{eq:grid}), and predictions outside this range would fall apart. This issue could of course be addressed by increasing the range of the grid, but to maintain the same grid density would mean that the number of training points increases by the order of $\mathcal{O}(n^p)$, where $p$ is the number of ODE parameters, thereby heavily increasing the computational cost. A data-dependent approach such as that described in \cite{ramsay2007parameter} is much more generalized. \\

Our methods naturally inherit all the challenges in utilizing ANNs. In our case, applying a $z$-standardization was sufficient, but for certain dynamical systems, the numerical solutions can have a few extremely large or small values, rendering the training targets to be rather skewed in distribution. With no modifications, the training of the ANN could easily neglect the fit around such values. This can, however, be circumvented by performing transformations on the training examples such as quantile transforms where necessary, which may be investigated in future research.

\section{Conclusion}
\label{section:conclusion}

Artificial neural networks, and deep learners in general are versatile and powerful function approximation tools that can be incorporated into widely-used Bayesian inference frameworks. We have demonstrated that the performance of our hybrid Bayesian-ANN computational framework is comparable to MCMC methods, and has the ability to circumvent certain shortcomings of such methods in estimating ODE parameters - mainly posterior intractability and computational efficiency issues. Regularization of the ANN using appropriate derivative penalties proved to be rather crucial in improving posterior inference due to the complexity of the solution surface of ODEs. \\

The strengths and limitations of our methods give rise to the following research avenues to be explored in future work. The ANN approach can be modified and extended to a sequential data assimilation setting to carry out inference on dynamic ODE parameters. A data-dependent method that involves ANN may also be explored to overcome the restrictions arising from the design of our methods. Since these research areas are tightly connected, they may be carried out in conjunction by devising a data-dependent sequential algorithm that utilizes ANNs to draw inference on dynamic ODE parameters. One may also extend the current ANN architecture to high dimensional settings to obtain further improvements on the computational costs over long run iterative methods.

\section*{Acknowledgement}

Funding: This research is mainly funded by an FRGS grant (FRGS/1/2020/STG06/HWUM/02/1) from the Ministry of Higher Education, Malaysia, and supported by the school of Mathematical and Computer Sciences, Heriot-Watt University (https://www.hw.ac.uk/). The funders had no role in the study design, data collection and analysis, decision to publish, or preparation of the manuscript. We also wish to extend our gratitude to Prof. Gavin J. Gibson for his contributions to this article.

\Urlmuskip=0mu plus 1mu\relax
\bibliographystyle{vancouver}
\bibliography{bibfile}

\end{document}